\newcommand{\DocVersion}{Accepted by The Astronomical Journal -- 15 Aug. 2012}

\pdfoutput=1

\documentclass[11pt,preprint]{aastex}

\slugcomment{\DocVersion}

\shortauthors{Shuping~et~al.}
\shorttitle{Spectral classification of bright sources in W40.}

\newcommand{\msun}{M$_{\sun}$}

\newcommand{\wforty}{\object[W 40]{W40}}

\newcommand{\irsonea}{\object[NAME W 40 IRS 1A]{IRS~1A}}
\newcommand{\irsoneanorth}{IRS~1A~North}
\newcommand{\irsoneasouth}{IRS~1A~South}
\newcommand{\irsoneb}{\object[NAME W 40 IRS 1b]{IRS~1B}}
\newcommand{\irsonec}{\object[NAME W 40 IRS 1c]{IRS~1C}}
\newcommand{\irsoned}{\object[NAME W 40 IRS 1d]{IRS~1D}}
\newcommand{\irsonee}{IRS~1E}
\newcommand{\irsonef}{IRS~1F}

\newcommand{\irstwoa}{\object[NAME W 40 IRS 2a]{IRS~2A}}
\newcommand{\irstwob}{\object[NAME W 40 IRS 2b]{IRS~2B}}
\newcommand{\irstwoc}{\object[NAME W 40 IRS 2c]{IRS~2C}}
\newcommand{\irstwod}{IRS~2D}
\newcommand{\irstwoe}{IRS~2E}
\newcommand{\irstwof}{IRS~2F}

\newcommand{\irsthreea}{\object[NAME W 40 IRS 3a]{IRS~3A}}

\newcommand{\irsfive}{IRS~5}

\begin{document}

\title{Spectral classification of the brightest objects in the galactic star forming region W40}

\author{R. Y. Shuping\altaffilmark{1,2}}
\affil{Space Science Institute \\
4750 Walnut Street \\
Suite 205 \\
Boulder, CO 80301}
\email{rshuping@spacescience.org}

\and

\author{William D. Vacca\altaffilmark{2}}
\affil{Universities Space Research Assoc.\\
Stratospheric Observatory for Infrared Astronomy \\
NASA Ames Research Center \\
MS 211-3 \\
Moffett Field, CA  94035
}

\and

\author{Marc Kassis}
\affil{
W. M. Keck Observatory \\
65-1120 Mamalahoa Hwy. \\
Kamuela, HI 96743
}

\and

\author{Ka Chun Yu\altaffilmark{2}}
\affil{
Denver Museum of Nature \& Science \\
2001 Colorado \\
 Denver, CO 80205-5798
}

\altaffiltext{1}{Universities Space Research Assoc.}
\altaffiltext{2}{Visiting Astronomer at the Infrared Telescope Facility, which is operated by the University of Hawaii under Cooperative Agreement no. NNX-08AE38A with the National Aeronautics and Space Administration, Science Mission Directorate, Planetary Astronomy Program.}

\begin{abstract}

We present high S/N, moderate resolution ($R\approx2000$) near-infrared spectra, as well as 10~\micron\ imaging, for the brightest members of the central stellar cluster in the W40 \ion{H}{2} region, obtained using the SpeX and MIRSI instruments at NASA's Infrared Telescope Facility (IRTF). Using these observations combined with archival Spitzer Space Telescope data, we have determined the spectral classifications, extinction, distances, and spectral energy distributions for the brightest members of the cluster.   Of the eight objects observed, we identify four main sequence (MS) OB stars (one late-O, 3 early-B), two Herbig Ae/Be stars, and two low-mass young stellar objects (Class~II).  Strong \ion{He}{1} absorption at 1.083~\micron\ in the MS star spectra strongly suggests that at least some of these sources are in fact close binaries.  Two out of the four MS stars also show significant infrared excesses typical of circumstellar disks.  Extinctions and distances were determined for each MS star by fitting model stellar atmospheres to the SEDs.  We estimate a distance to the cluster of between 455 and 535~pc, which agrees well with earlier (but far less precise) distance estimates.  We conclude that the late-O star we identify is the dominant source of LyC luminosity needed to power the W40 \ion{H}{2} region and is the likely source of the stellar wind that has blown a large ($\approx 4$~pc) pinched-waist bubble observed in wide field mid-IR images.  We also suggest that 3.6~cm radio emission observed from some of the sources in the cluster is likely not due to emission from ultra-compact \ion{H}{2} regions, as suggested in other work, due to size constraints based on our derived distance to the cluster.  Finally, we also present a discussion of the curious source \irsthreea, which has a very strong mid-IR excess (despite its B3 MS classification) and appears to be embedded in a dusty envelope roughly 2700~AU in size.

\end{abstract}



\section{Introduction}
\label{Intro}

\wforty\ is a thermal radio continuum source associated with the molecular cloud G~28.74+3.52 in the Aquila Rift~\citep{Westerhout:1958fk,Zeilik:1978uq}. A faint, patchy optical \ion{H}{2} region (Sh~64) is  associated with the radio emission~\citep{Sharpless:1959}.   Radio observations imply a total Lyman continuum luminosity of $\sim1.5\times10^{48}$ photons~s$^{-1}$~\citep{Goss:1970vn,Smith:1985}---about 10\% of the ionizing flux found in the Trapezium Cluster in Orion---corresponding to  several early B stars or a single O9V.  An IRAS source, cold ammonia core, and a number of mm-wave sources are also associated with the cluster~\citep{Molinari:1996dz,Maury:2011}, suggesting on-going star formation from dense molecular material.  In addition, \citet{Crutcher:1982wd} and \citet{Wu:1996} have found multiple CO line components and/or extended line wings toward \wforty\ which suggests outflows driven by young stellar objects (YSOs).

Early infrared maps of the central 3\arcmin\ of \wforty\ revealed 7 bright 
sources with optical counterparts behind $9 - 10$ magnitudes of visual extinction~\citep{Crutcher:1982wd,Smith:1985}.  Spectral energy distributions (SEDs) from IR through millimeter wavelengths suggest substantial circumstellar material around most of these bright sources~\citep{Smith:1985,Vallee:1994ai}.  2MASS images of the region reveal a cluster of near-IR sources within the central 5\arcmin\ of the brightest sources (Fig.~\ref{fig:W40_2MASS_K_cluster}).  Recent radio observations at 3.6~cm show a cluster of compact sources in the central portion of the W40 IR cluster, many of which correspond to the known infrared sources~\citep{Rodriguez:2010}.  X-Ray observations of the W40 cluster with the Chandra X-Ray Observatory (CXO) reveal approximately 200 sources associated with the cluster,  the majority of which are thought to be low-mass young stellar objects~\citep{Kuhn:2010}.  TheX-Ray luminosity function (XLF) obtained by the Chandra  Orion Ultra-deep Project (COUP) indicates a completeness limit of 70\% (down to 0.1~\msun) and suggests a total cluster population of $\approx$600 objects~\citep{Kuhn:2010}.  UKIRT near-IR observations presented in the same study show that a significant fraction of the X-ray sources also have K-band excesses indicative of circumstellar disks and suggest an age of $< 1$ Myr~\citep{Kuhn:2010}.  {\it Herschel} SPIRE and PACS far-IR maps of the entire Aquila Rift region (which includes the Serpens star forming regions and MWC297/Sh-62) reveal $\approx$201 sources, most of which appear to be associated with W40~\citep{Bontemps:2010}.  Analysis of the far-IR colors suggest that 45--60 of the detected sources are deeply embedded Class~0 objects~\citep{Bontemps:2010}. 

Depending on the technique used, the distance to W40 ranges from 300 to 900 pc~\citep[and references therein]{Rodney:2008}, making \wforty\ one of  only a few  \ion{H}{2} regions within a kiloparsec of the Sun.  XLF fitting by \citet{Kuhn:2010} yields a best-fit distance of $\approx 600$~pc, while \citet{Bontemps:2010} adopt a much closer distance of 260~pc based on colocation of W40 with the Serpens region on the sky.  The distance to W40 will be discussed further in Section~\ref{distance}.  

Many questions remain regarding the initial mass function (IMF), age, and star
formation history of W40. Even the brightest members of the cluster have not been studied in any detail. In this paper we present infrared observations of these bright members of the W40 IR cluster in an effort to better constrain some of their basic properties, including spectral type, age, and distance.   This is just a first step in better understanding the cluster and star forming region as a whole.

\section{Source Identification}
\label{source_ids}

\begin{figure}
\centering
\epsscale{0.5}
\plotone{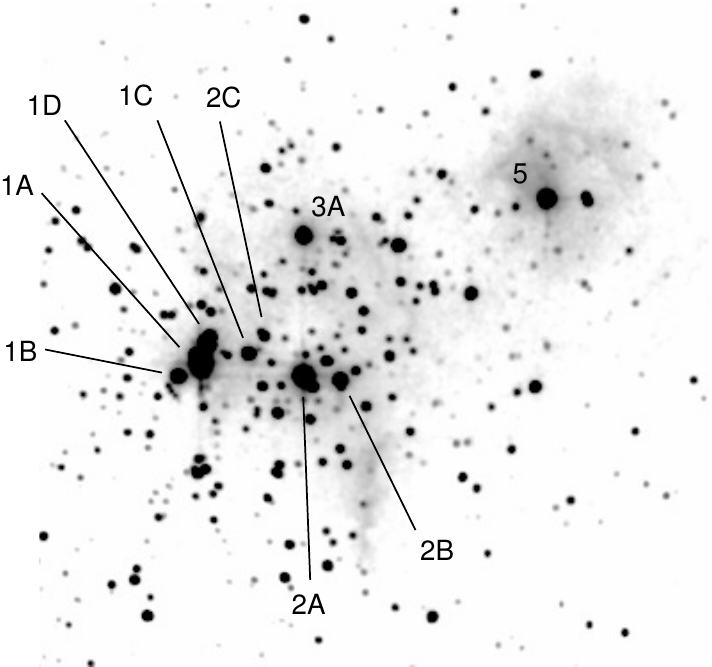}
\caption{2MASS K-band image of the W40 IR stellar cluster, approximately 6\arcmin\ field of view (North is up, East to the left).  Source designations are from \citet{Smith:1985} except for \irsfive\ which has not been previously reported.  Many of the original IR sources reported by \citet{Smith:1985} are revealed to be multiples. (This mosaic was created using Montage v3.0:  Montage is maintained by the NASA Infrared Science Archive (IRSA), part of the Infrared Processing and Analysis Center (IPAC)).}
\label{fig:W40_2MASS_K_cluster}
\end{figure}

In Figure~\ref{fig:W40_2MASS_K_cluster} we present the 2MASS K-band image of the central 6\arcmin\ of the cluster with the brightest sources identified and labelled using the \citet{Smith:1985} naming convention.  Inspection of 2MASS K-band image reveals that some of the original IR sources detected by \citet{Smith:1985} are in fact multiples.  In addition, there is another bright K-band source approximately 2\arcmin\ to the west of \irsthreea\ that corresponds to a bright star on the DSS R plate but is not reported by either \citet{Zeilik:1978uq} or \citet{Smith:1985}.  In keeping with the current naming convention, we will refer to this source as IR Source 5 (Optical source 4A was not detected in the IR by \citet{Smith:1985}).  \irsfive\ is surrounded by patchy nebulosity on both the DSS R plate and in the 2MASS K images.  

\section{Observations}
\label{observations}

\subsection{SpeX Observations}
\label{sect:spex_obs}

Observations of the central portion of the W40 IR cluster were carried out using NASA's Infrared Telescope Facility (IRTF) with the SpeX instrument in  2006 and then again in 2010.   SpeX is the facility near-IR spectrograph for the IRTF and provides spectral resolving power ($R$) of 1000 -- 2500 across 0.8--2.4,  2.0--4.1, and 2.3--5.5~\micron , using prism cross-dispersers with a Raytheon Aladdin 3 1024$\times$1024 InSb array~\citep{Rayner:2003}.  SpeX also contains an infrared slit-viewer/guider covering a 60$\times$60\arcsec\ field-of-view with a Raytheon Aladdin 2 512$\times$512 InSb array.

\begin{figure}
\epsscale{0.8}
\plotone{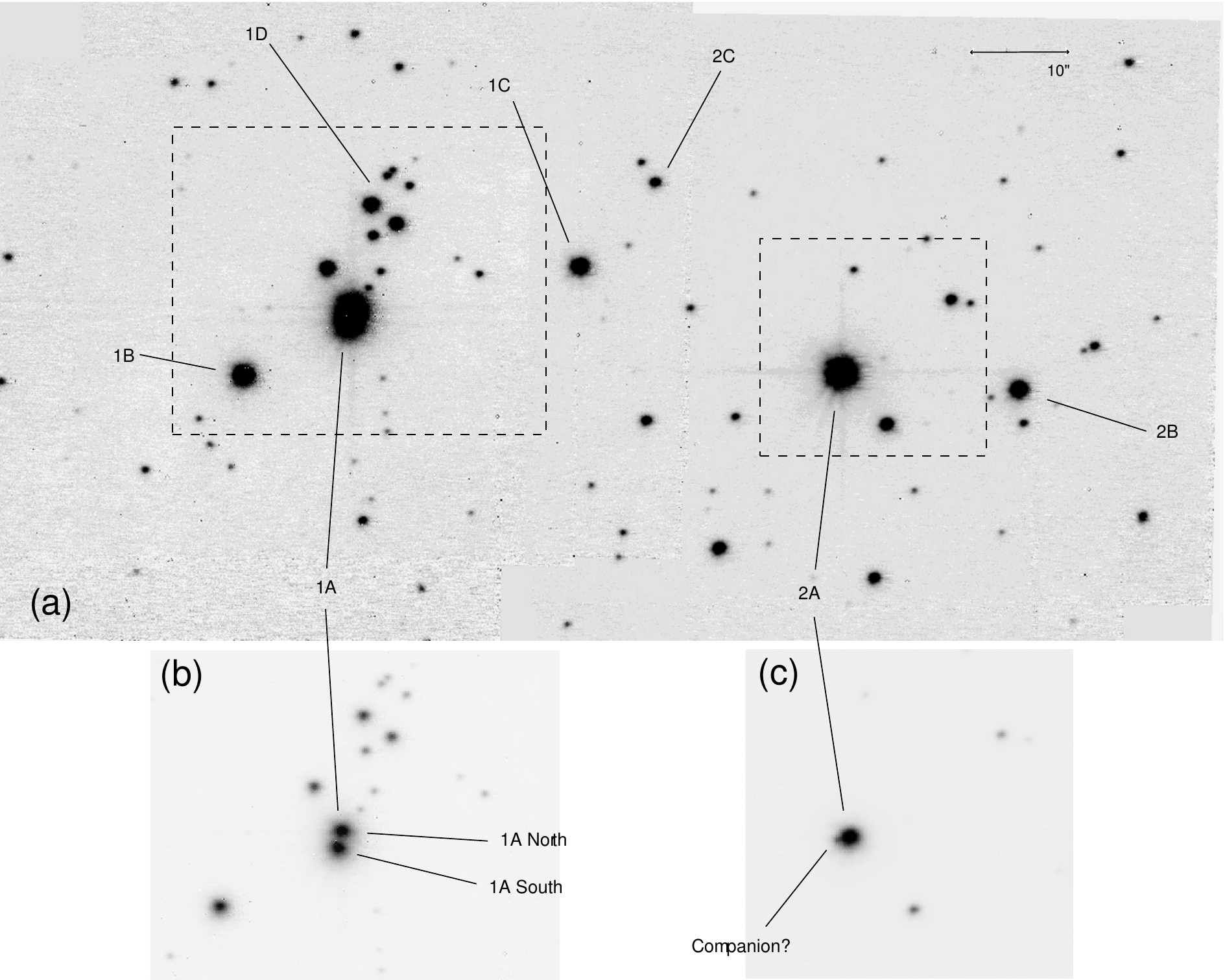}
\caption{(a) SpeX guide camera K-band image of the region surrounding IRS 1 and 2.  (b) Detail of IRS A showing resolution of 1A into North and South components. (c ) Detail of IRS 2A showing possible companion to the east.}
\label{fig:IRS12_GuideDog_K}
\end{figure}

We obtained both images and spectra of the central region of W40 on 8~July~2006 under clear skies with K-band seeing of 0$\farcs$4--0$\farcs$6~(FWHM).  Images of the central region of W40 were taken using the SpeX guide camera with a 9-point dither pattern centered on IRS 1A and 2A respectively.  Dithered image sets were reduced using standard techniques and then mosaiced together; the final mosiaced image is shown in Figure~\ref{fig:IRS12_GuideDog_K}.  \irsonea\ is clearly resolved into two bright sources (``North'' and ``South''), separated by $\approx 2$\arcsec\ along a roughly N-S position angle, and 3 additional weaker sources to the NE and NW.  \irsoned\ is resolved into a small cluster of at least 7 distinct sources.  IRS 2A appears to have a faint companion to the East as well (see Figure inset), but higher resolution observations are required for confirmation.  

Moderate resolution spectra (R$\approx$2000) were  obtained for each of the brightest near-IR sources (Table~\ref{table:obs_summary}).  The observations were acquired in ``pair mode", in which the object was observed at two separate positions (``A'' and ``B'') along the 15\arcsec-long slit. The slit width was set to 0$\farcs$3, which yields a nominal resolving power of 2000 for the short-wavelength cross-dispersed (SXD) spectra and 2500 for the long-wavelength cross-dispersed (LXD2.1) spectra. The slit was set to the parallactic angle during the observations.   In SXD mode, we used exposure times of 60~s for the brightest sources and 120~s for the fainter sources and cycled through the AB pairs twice yielding total on-source exposure times of 240 and 480~s respectively.  This mode yields spectra spanning the wavelength range 0.8 -2.5~\micron\ divided into 6 spectral orders.   In LXD mode, we observed only the brightest targets with 30~s integration times, cycling through the AB pairs twice for \irsoneb\ and 2A, and 6 times for \irsthreea . This mode yields spectra covering 2.2$-$5.1~\micron\ in 6 spectral orders.  The airmass was 1.1--1.6 for all SXD observations and 2.0--2.3 for the LXD observations. The A0~V stars HD~171149 and HD~169961 were observed throughout the night as telluric and flux calibration standards. The airmass difference between the objects and the standards was always $< 0.2$.  A set of internal flat fields and arc frames were obtained immediately after each set of observations of W40 for flat fielding and wavelength calibration purposes.

\begin{deluxetable}{lccccccccc}
\tabletypesize{\tiny}
\tablewidth{0pt}
\tablecaption{Sources Observed with SpeX}
\tablehead{
\colhead{Source}  &  \colhead{2MASS}  &  
\colhead{2MASS} & \colhead{2MASS} &
  &  &  & \colhead{Exp. Time (s)} &  \colhead{Exp. Time (s)}  \\
 &  \colhead{ID}  &  
\colhead{RA (J2000)} & \colhead{Dec (J2000)} &
\colhead{J}  & \colhead{H}  &  \colhead{K} &
\colhead{(SXD)}  &  \colhead{(LXD)} 
}
\startdata					
IRS 1A N\tablenotemark{a} &  18312782-0205228  &  18h31m27.81s  &  -02d05m21.87s  &  9.31\tablenotemark{a}  &  7.49\tablenotemark{a}  &  6.13\tablenotemark{a} & 900\tablenotemark{b} & 300\tablenotemark{c}  \\
IRS 1A S \tablenotemark{a} &  18312782-0205228  &  18h31m27.83s  &  -02d05m23.61s  &  8.11\tablenotemark{a}  &  7.00\tablenotemark{a}  &  6.49\tablenotemark{a} & 480 & 300\tablenotemark{c} \\
\irsoneb  &  18312866-0205297  &  18h31m28.66s  &  -02d05m29.78s  &  11.47  &  9.15  &  7.50 & 480 &  120  \\
\irsonec  &  18312601-0205169  &  18h31m26.02s  &  -02d05m16.97s  &  10.87  &  9.11  &  8.08  & 480 &  \\
\irstwoa  &  18312397-0205295  &  18h31m23.97s  &  -02d05m29.51s  &  8.74  &  7.19  &  6.00 & 240 & 120    \\
\irstwob  &  18312257-0205315  &  18h31m22.58s  &  -02d05m31.60s  &  9.72  &  8.79  &  8.29 & 480 &   \\
\irsthreea  &  18312395-0204107  &  18h31m23.95s  &  -02d04m10.76s  &  8.91  &  8.02  &  7.46 & 240 & 360   \\
\irsfive  &  18311482-0203497  &  18h31m14.82s  &  -02d03m49.80s  &  8.27  &  7.51  &  7.08  & 240 &  \\
\enddata
\tablecomments{All SpeX observations carried out 08~July~2006 (UT) except where noted.  JHK photometry is from 2MASS except where noted; relative uncertainty for the 2MASS JHK photometry is less than 2 \%.  }
\tablenotetext{a}{\irsonea\ is not resolved in the 2MASS images.  Positions listed for \irsoneanorth\ and South are derived from our SpeX Guide Dog images; photometry is derived from the SpeX JHK spectra with a relative uncertainty of roughly 10 \%.}
\tablenotetext{b}{Observed 09 July 2010 (UT).}
\tablenotetext{c}{Observed 13 May 2010 (UT).}
\label{table:obs_summary}
\end{deluxetable}

In order to improve our S/N, we re-observed \irsonea\ North and South on 13~May and 08--09 July 2010.  Conditions for the May observations were good:  clear skies and $\sim 0\farcs7$ seeing in K.  In July conditions were excellent, with photometric skies and seeing from 0$\farcs$3 -- 0$\farcs$7 in K.  Observational procedures were the same as for the 2006 observations except with longer integration times and additional pair cycles to improve S/N.  Calibration and reductions were also identical to the 2006 observation run.  Note, however, that the S/N for \irsoneanorth\ in the LXD mode was limited by the calibration source which was significantly fainter than \irsoneanorth .  

\begin{figure}
\epsscale{0.49}
\plotone{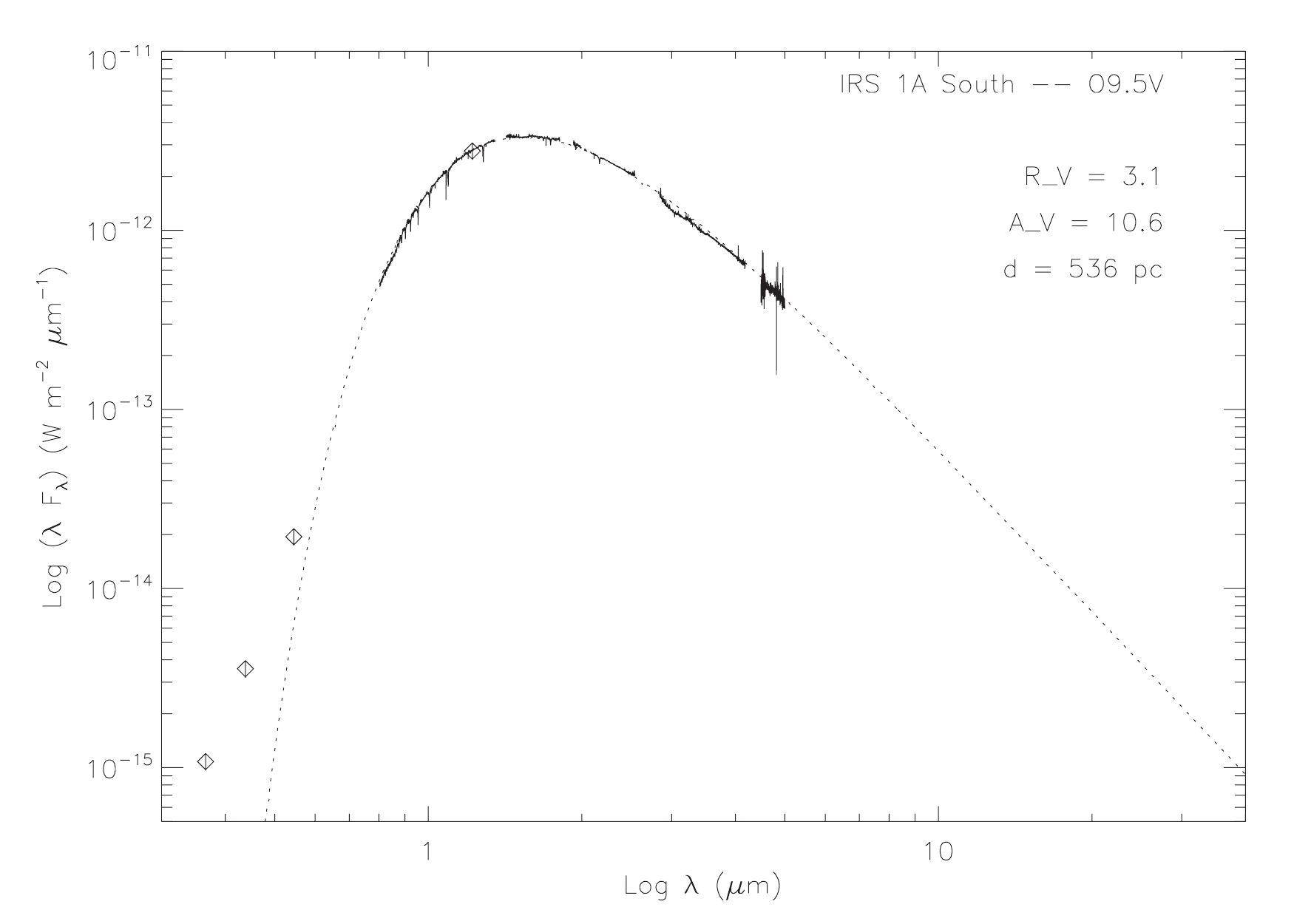}
\plotone{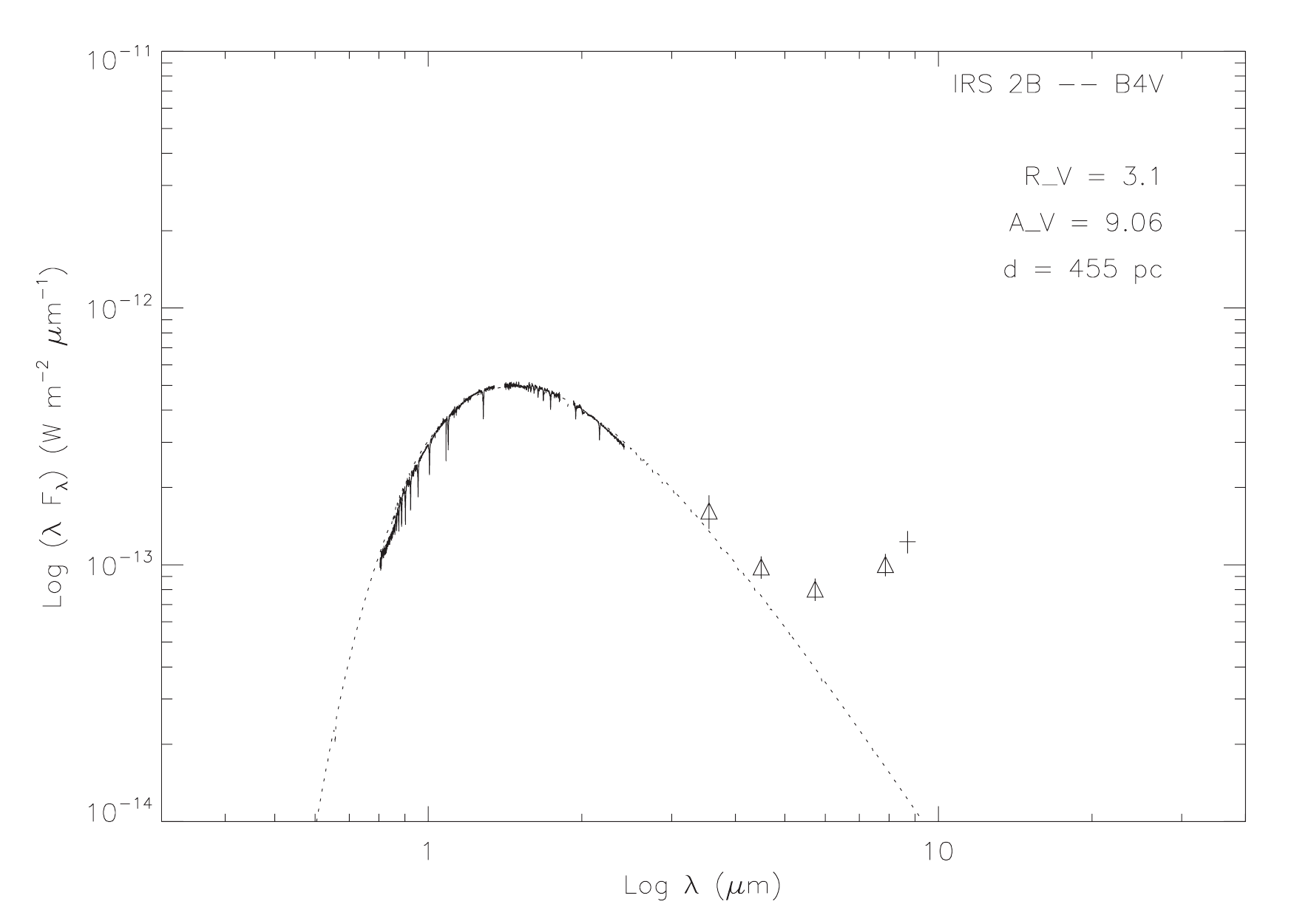}
\plotone{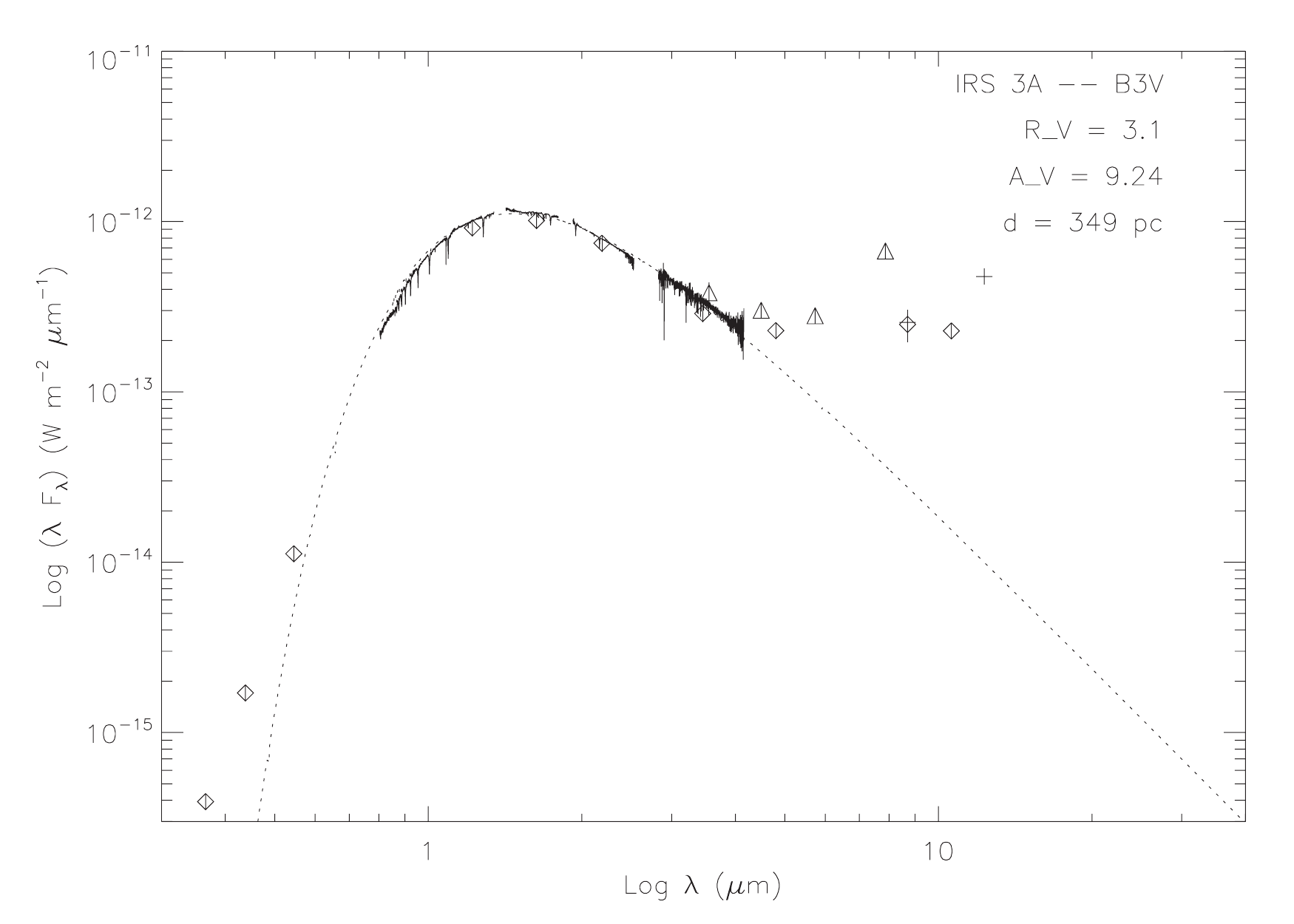}
\plotone{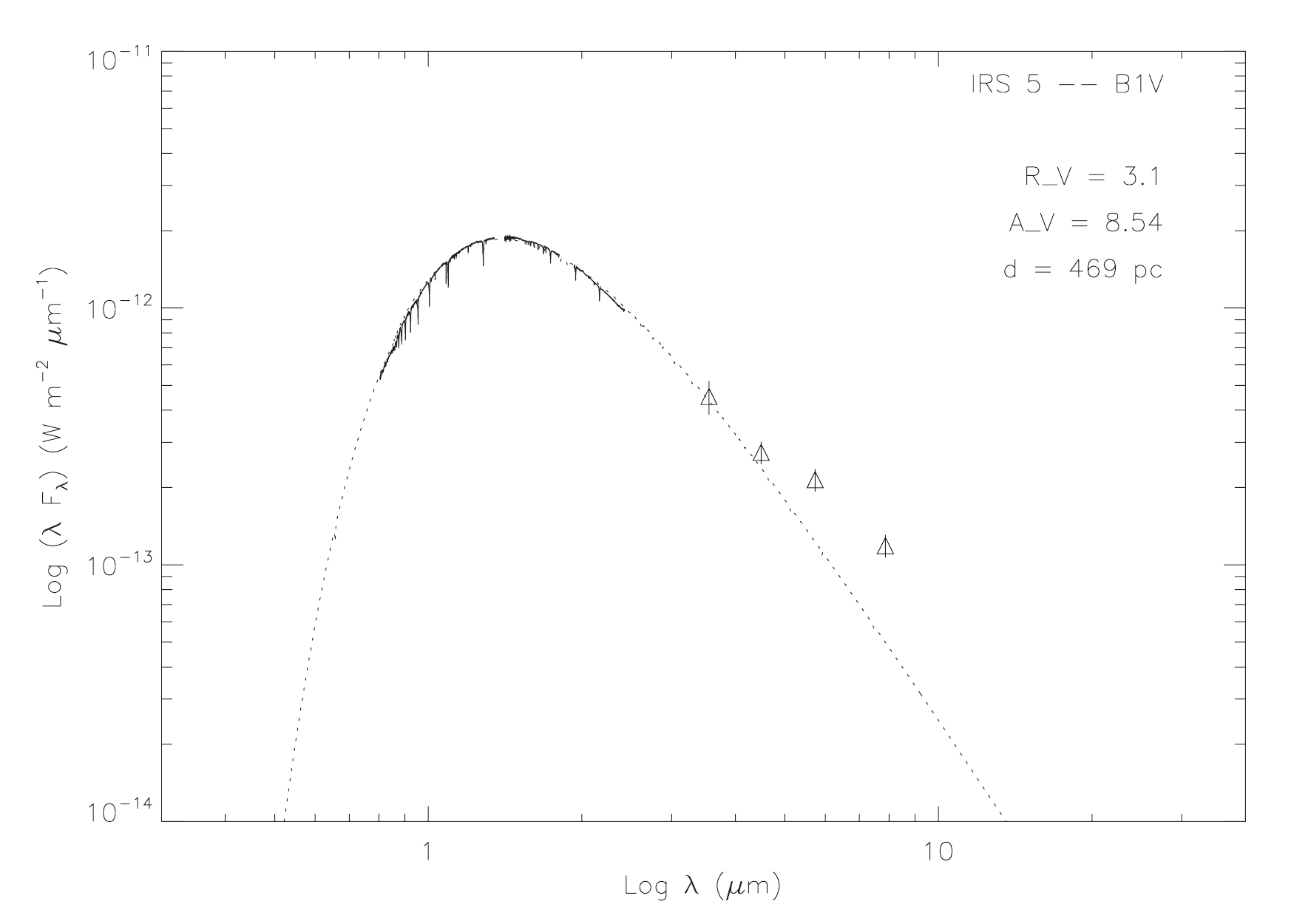}
\caption{Spectral energy distributions for main sequence sources in W40.  Solid line---SpeX near-IR spectra (this study); Crosses---MIRSI mid-IR photometry (this study); Triangles---Spitzer-IRAC photometry (this study);  Diamonds---\citet{Smith:1985} photometry; Dotted line---stellar atmospheric model fit for given reddening and distance (see text).}
\label{fig:SEDs_MS}
\end{figure}

\begin{figure}
\epsscale{0.49}
\plotone{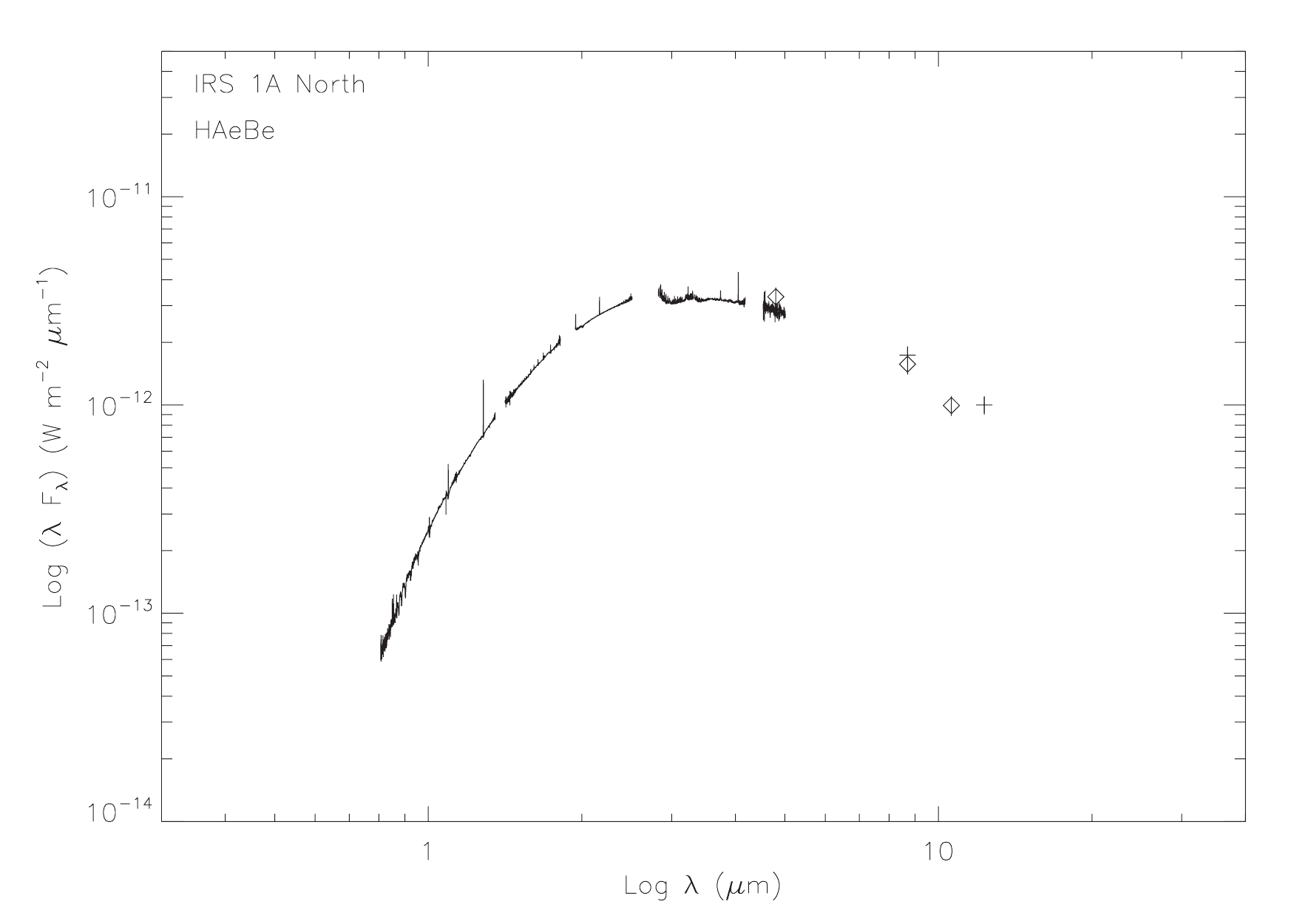}
\plotone{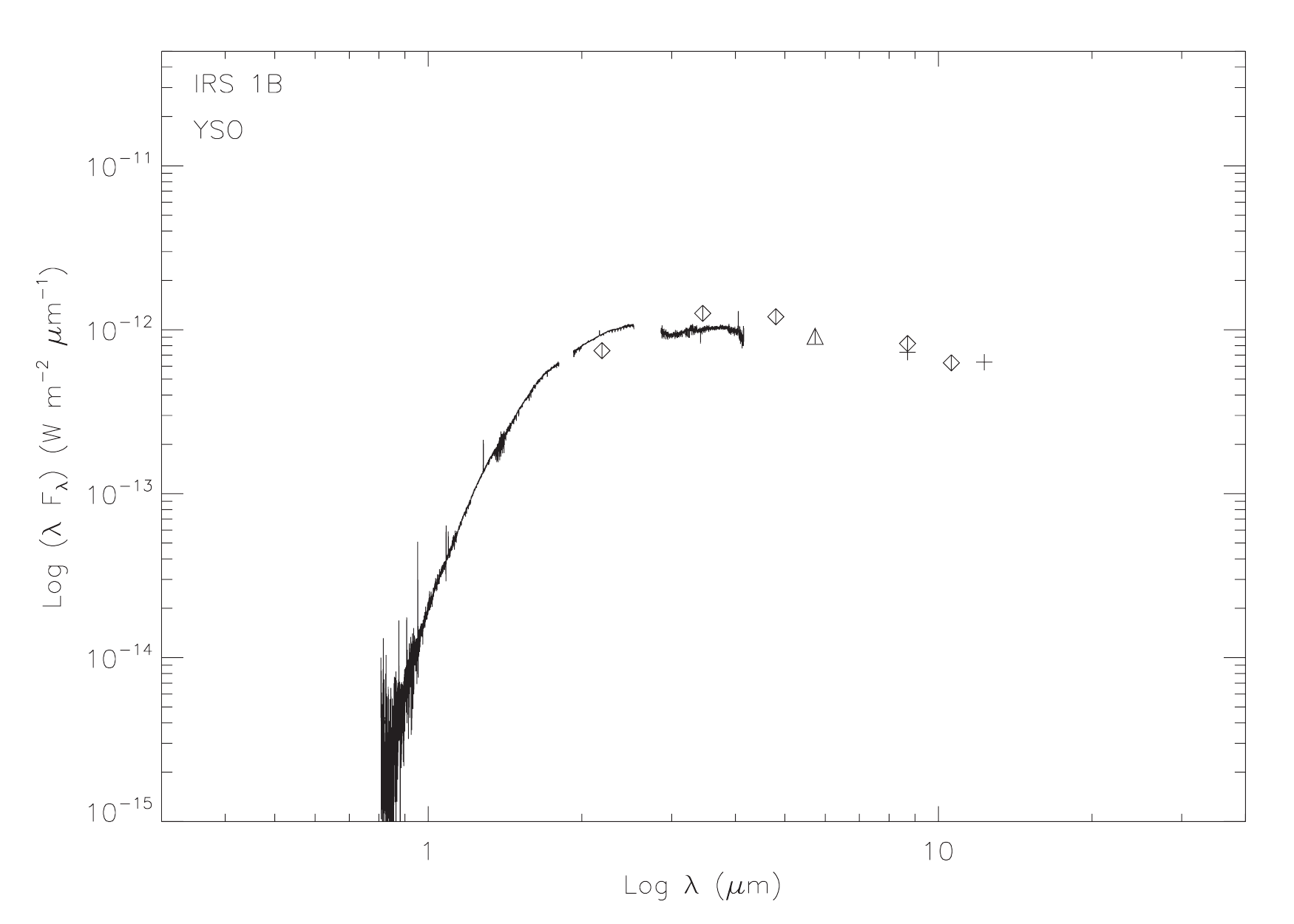}
\plotone{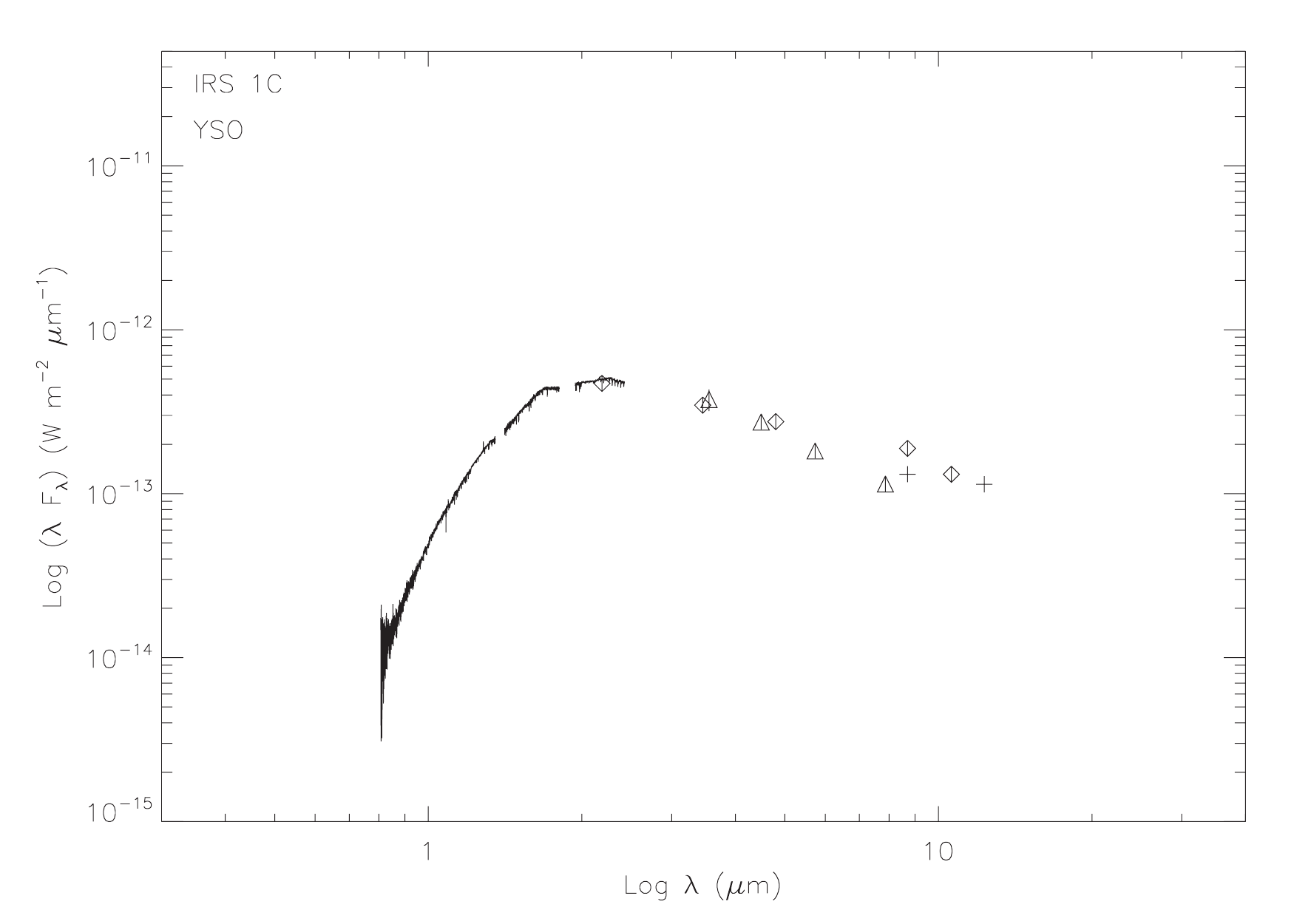}
\plotone{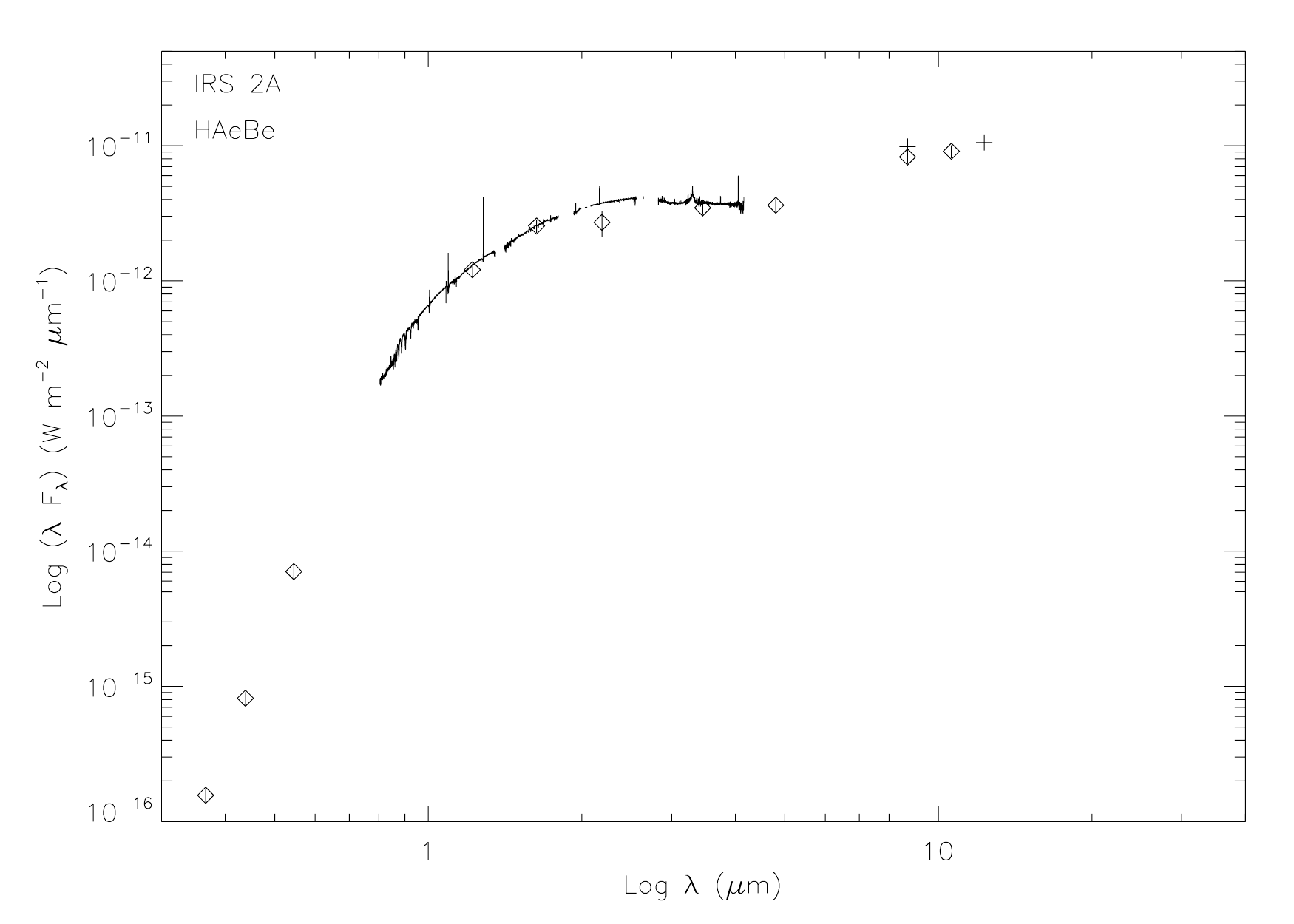}
\caption{Spectral energy distributions for young stellar objects and pre-main sequence stars in W40.  Line---SpeX near-IR spectra (this study); Crosses---MIRSI photometry (this study); Triangles---Spitzer-IRAC photometry (this study);  Diamonds---\citet{Smith:1985} photometry.}
\label{fig:SEDs_YSOs}
\end{figure}

All data were reduced using Spextool \citep{Cushing:2004}, the IDL-based package developed for the reduction of SpeX data. The Spextool package performs non-linearity corrections, flat fielding, image pair subtraction, aperture definition, optimal extraction, and wavelength calibration. The sets of spectra resulting from the individual exposures were median combined and then corrected for telluric absorption and flux calibrated using the extracted A0\, V telluric standard spectra and the technique and software described by \citet{Vacca:2003}.  Uncertainty in the SpeX flux calibration is $\sim 10$ \% \citep{Cushing:2005,Rayner:2009}.  The spectra from the individual orders were then spliced together by matching the flux levels in the overlapping wavelength regions, and regions of poor atmospheric transmission were removed. The complete, flux calibrated  $0.8-5.0$~\micron\ spectra are shown in Figures~\ref{fig:SEDs_MS} and \ref{fig:SEDs_YSOs}.  

\subsection{MIRSI Observations}

Mid-IR observations of the central portion of the W40 IR cluster were carried out at the IRTF using the Mid-InfraRed Spectrometer and Imager~\citep[MIRSI,][]{Kassis:2008} instrument  on 8 July 2006 (UT).  MIRSI offers complete spectral coverage over
the atmospheric windows at 8-14~\micron\ and 18-26~\micron\ for both imaging
(discrete filters and a circular variable filter) and spectroscopy (in
the 10 and 20~\micron\ windows with resolutions of 200 and 100,
respectively) utilizing a Raytheon 320$\times$240 pixel Si:As  array with a plate scale of 0$\farcs$27 per pixel, and a field-of-view of 86\arcsec$\times$63\arcsec\ at the IRTF. Ê

\begin{figure}
\epsscale{1.0}
\plottwo{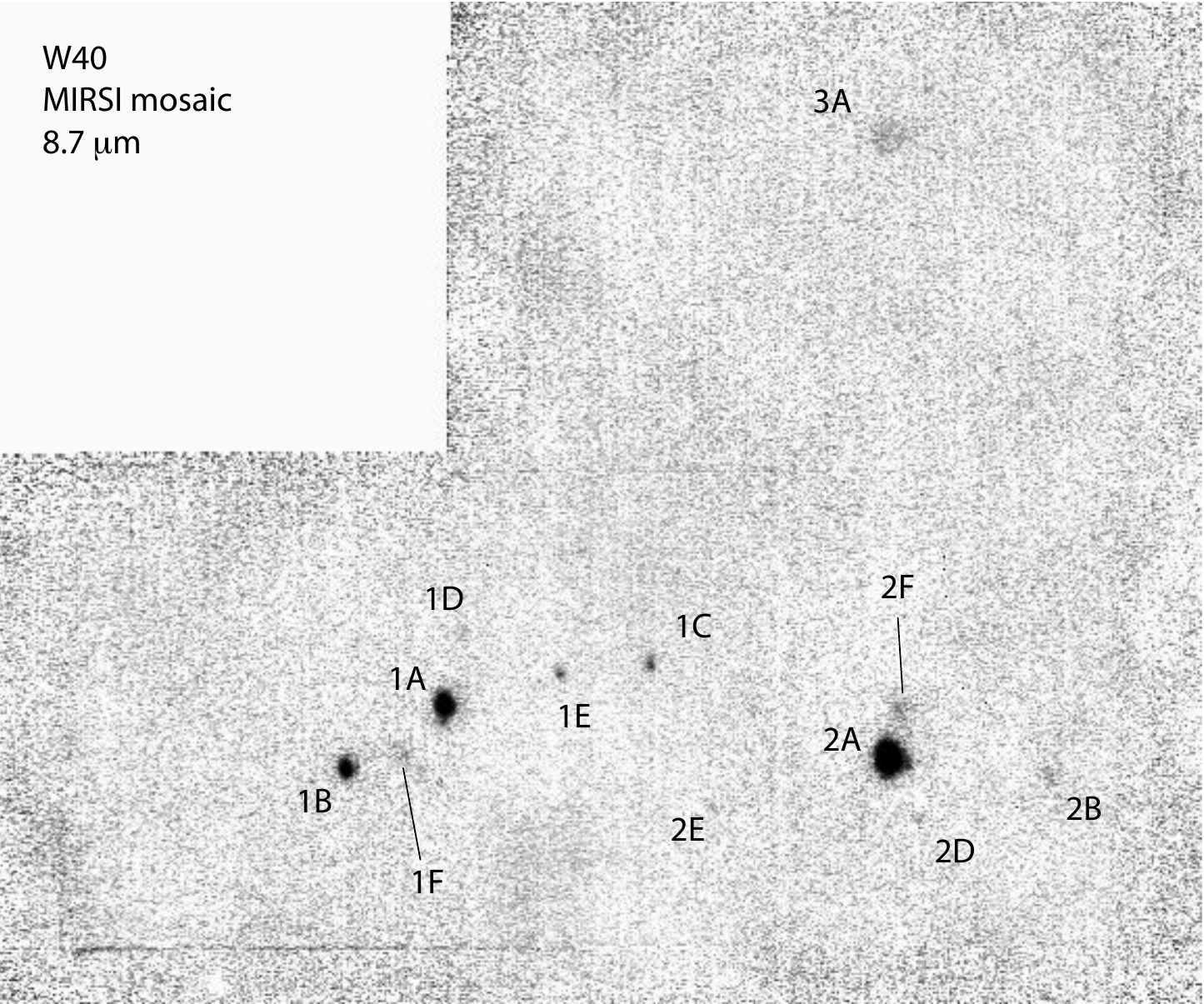}{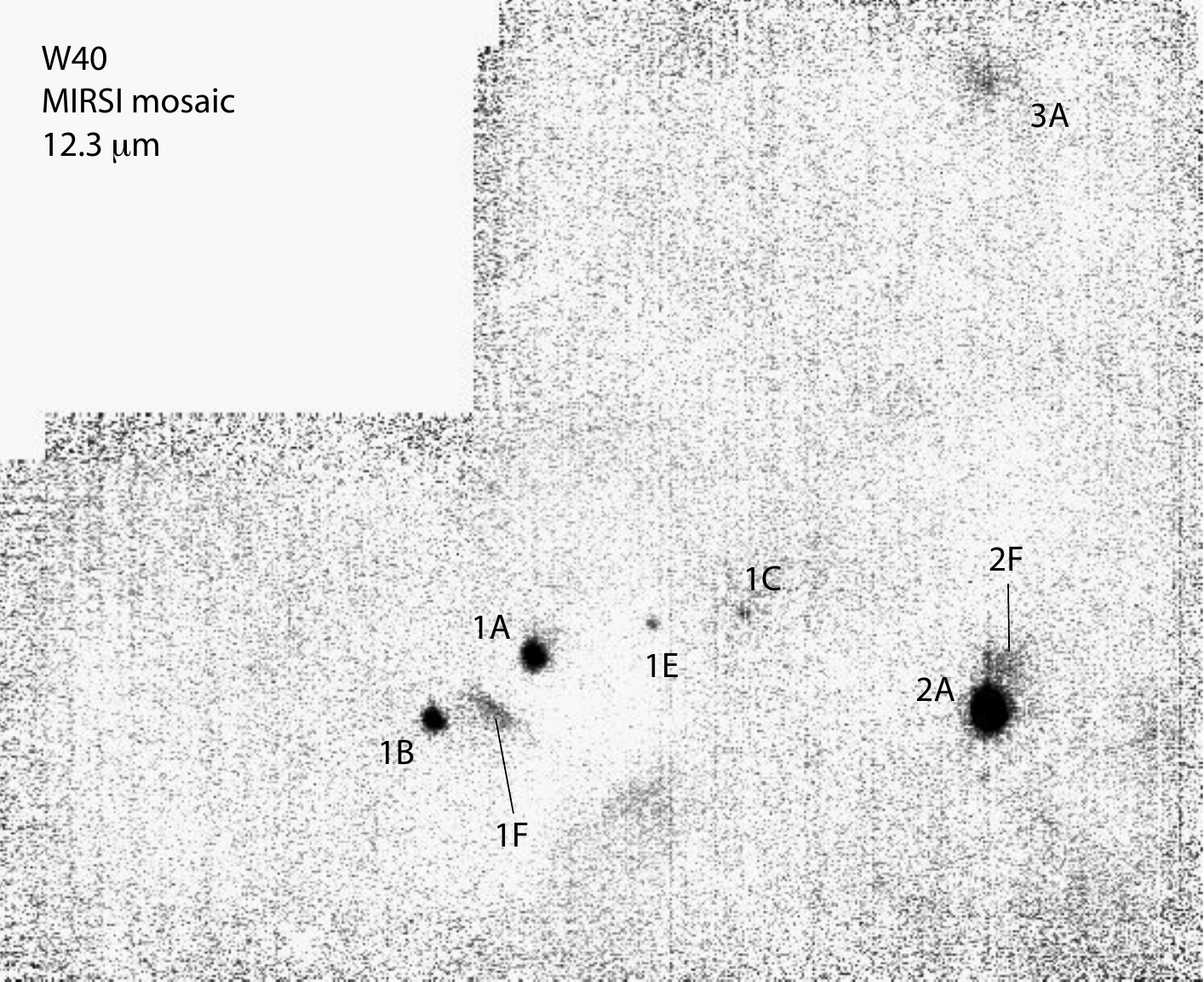} 
\caption{MIRSI mosaic of W40 IRS 1 -- 3 at 8.7 and 12.3~\micron.  Final mosaic is 2.6\arcmin $\times$ 2.3\arcmin (North is up, East is left).}
\label{fig:W40_MIRSI}
\end{figure}

Images of the central 3\arcmin\ of the IR cluster were obtained using the 8.7 and 12.3~\micron\ filters.  Approximately 15 pointings were used to cover IRS 1 through 3 and IRS 5 with on-source exposure times of 25~s for each pointing in each filter.  Standard chop and nod techniques were used for each pointing to eliminate infrared background emission.  Standard stars $\alpha$ Lyr (A0V) and $\mu$ Cep (M2Ia) were observed throughout the night for flux calibration.  The final mosaics are shown in Figure~\ref{fig:W40_MIRSI}. Note that only the region surrounding IRS~1 -- 3 is shown as IRS 5 was not detected in any of the filters used.  The infrared sources 1E and F, and 2 D -- F are reported here for the first time and have been named according to the convention of \citet{Smith:1985}.   Fluxes derived using standard aperture photometry techniques are shown in Table~\ref{table:mirsi_photom}.  For the brighter sources, the uncertainty is dominated by systematic errors in the flux calibration, whereas for the fainter sources, it is dominated by uncertainty in the background emission.  

\begin{deluxetable}{lccl}
\tabletypesize{\scriptsize}
\tablewidth{0pt}
\tablecaption{MIRSI photometry}
\tablehead{
\colhead{Source}  &  
\colhead{8.7 \micron} &   
\colhead{12.3 \micron} &  
\colhead{Notes}   \\
  &  
\colhead{mJy (rel. err)} &   
\colhead{mJy (rel. err)}&  
}
\startdata					
\irsonea  & 5040.6 (0.06)  & 4104.2 (0.06)    &    \\
\irsoneb  &	2115.0 (0.1) &	2602.7 (0.1)  &  	 \\
\irsonec   &	381.3 (0.1) &	468.5 (0.1)  & 	 \\
\irsoned  &  176.6  (0.1)  &  $<$400     &  \\
\irsonee  &    289.8 (0.1)   &	 240.8 (0.1)    &  \\
\irsonef  &     364.9 (0.1)  &  1008.8  (0.1)   &  Diffuse \\
\irstwoa    &	28571.5 (0.06) &	43307.4 (0.06) &	17140.2 (0.2) \\	
\irstwob    &	357.1 (0.1) &	$<$400   & 	\\
\irstwoc    &	$<$300  &		$<$400   & Not detected	 	  \\
\irstwod  &  99.1  (0.1)  &   $<$400   	&  	 	\\
\irstwoe  &   255.7  (0.1)  &   $<$400   &		 	\\
\irstwof  &  394.4 (0.1)  &   241.1  (0.1)   & Diffuse  \\
\irsthreea  &	741.9 (0.1) &	1952.4 (0.1)  &  Resolved, non-point-like  \\		
\irsfive       &	$<$300  &		$<$400   	& Not detected	 	  \\
\enddata
\tablecomments{An aperture radius of 3\arcsec\ was used for all measurements.}
\label{table:mirsi_photom}
\end{deluxetable}


\clearpage

\section{Spitzer Space Telescope Archive Data}
\label{spitzer}

\begin{figure}
\epsscale{0.75}
\plotone{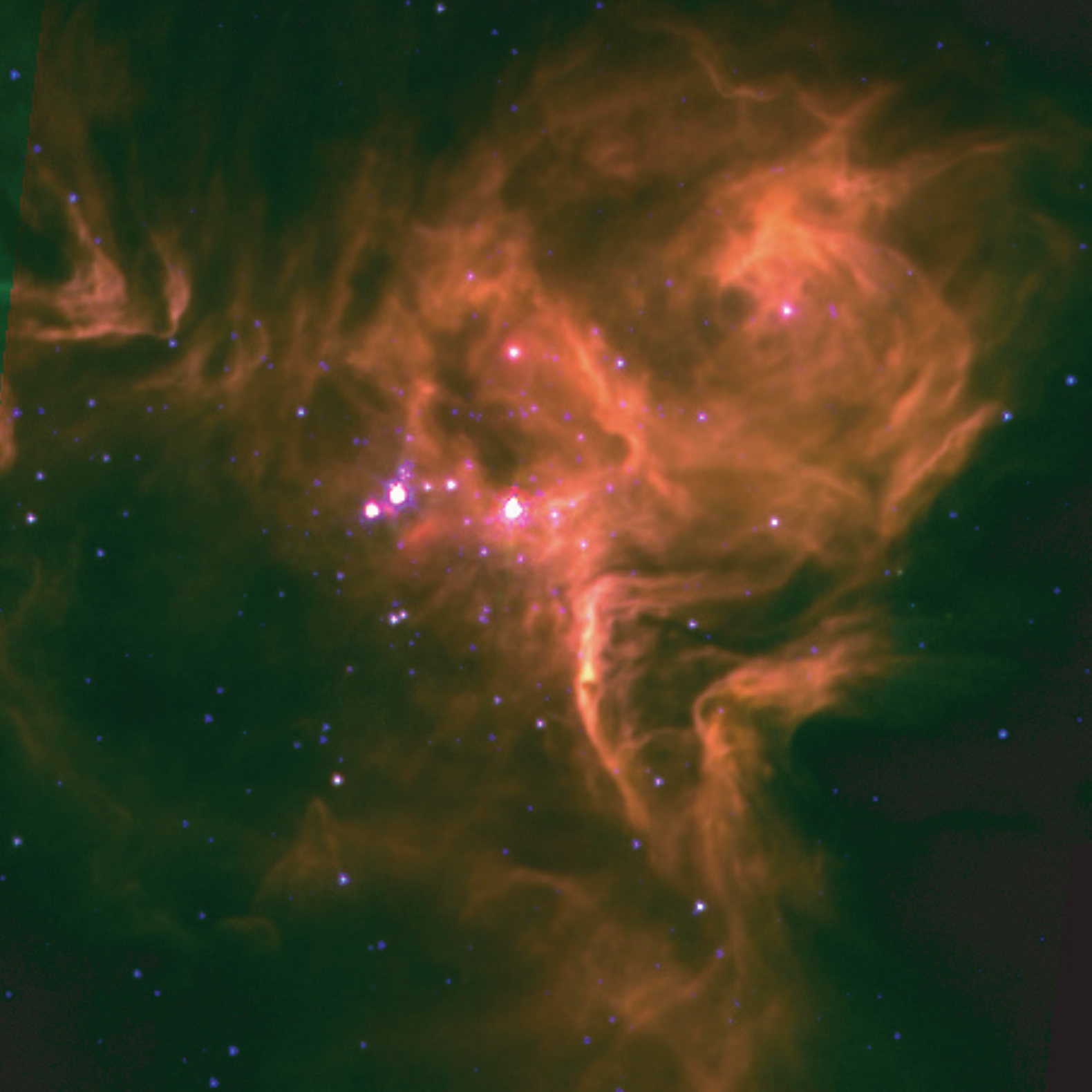}
\caption{Spitzer-IRAC composite mosaic of W40 at 8.0 (red), 5.8 (green), and 3.6~\micron\ (blue).  Field-of-view is roughly 10\arcmin\ (North is up, East is left).}
\label{fig:W40_IRAC_134}
\end{figure}

We also retrieved IRAC images of the W40 central cluster from the Spitzer Space Telescope Archive, which were originally obtained as part of Spitzer Program P30576 (Allen~{\it et al.}).  These observations were carried out in IRAC ``High Dynamic Range'' (HDR) mode so that a short (0.6~s) and long (12.0~s) exposure was obtained for each discrete pointing.  The central part of W40 suffers from numerous saturation effects/artifacts in the long exposure data, so we restricted our study to only the short exposure images.  We obtained the ``basic calibrated data'' (BCD) for all IRAC bands within the central 10\arcmin\ of \irstwoa\ using the \anchor{http://sha.ipac.caltech.edu/applications/Spitzer/SHA/}{Spitzer Heritage Archive}\footnote{
\url{http://sha.ipac.caltech.edu/applications/Spitzer/SHA/}}
 and then assembled a mosaic for each band (1, 2, 3, and 4) using \anchor{http://irsa.ipac.caltech.edu/data/SPITZER/docs/dataanalysistools/tools/mopex/}{MOPEX}\footnote{
 \url{http://irsa.ipac.caltech.edu/data/SPITZER/docs/dataanalysistools/tools/mopex/}}
~\citep{Makovoz:2005} with default settings.    A three-color composite image at 3.6, 5.8, and 8.0~\micron\ (channels 1, 3, and 4) is shown in Figure~\ref{fig:W40_IRAC_134}.  Simple aperture photometry was carried out to determine the fluxes of the brightest sources in each channel using the prescriptions and aperture corrections in the \anchor{http://irsa.ipac.caltech.edu/data/SPITZER/docs/irac/iracinstrumenthandbook/}{IRAC Instrument Handbook}\footnote{
\url{http://irsa.ipac.caltech.edu/data/SPITZER/docs/irac/iracinstrumenthandbook/}}.  
For all targets we picked the largest aperture possible that also did not contain any contaminating flux from the PRF wings of nearby sources. The appropriate aperture corrections were applied to all measurements. For IRS5 in Band 4, the background nebular emission varies substantially over small scales close the the star:  we chose a smaller aperture in this case to minimize contamination from the background nebula.   No color correction has been applied.  Fluxes and aperture sizes are listed in Table~\ref{table:irac_photom}.
According to the Handbook, errors should be $< $10\% for channels 2 --- 4 and $<$ 15\% for channel 1.  A few of the brightest sources are still saturated in some of the bands.  Since \irsthreea\ does not appear to be a point source (see Sect.~\ref{sect:irsthreea}), a large (12\arcsec) aperture was used with no correction.

\begin{deluxetable}{lccccc}
\tabletypesize{\scriptsize}
\tablewidth{0pt}
\tablecaption{IRAC Photometry}
\tablehead{
\colhead{Source}  &  
\colhead{3.6 \micron \tablenotemark{a}}  &  
\colhead{4.5 \micron \tablenotemark{b}} &
\colhead{5.8 \micron \tablenotemark{b}} & 
\colhead{8.0 \micron \tablenotemark{b}}  &
\colhead{Aperture}    \\
  &  
\colhead{mJy}  &  
\colhead{mJy} &
\colhead{mJy} & 
\colhead{mJy}  &
\colhead{Radius (\arcsec)\tablenotemark{c}}    \\
}
\startdata					
\irsonea &	$>$950 & $>$980 & $>$6950 & $>$3700 & SAT    \\
\irsoneb &	$>$950 &	$>$980 &	1742.4 &	$>$3700 &  6.0		 \\
\irsonec &	445.1 &	410.1 &	348.5 &	301.0 & 3.6		 \\
\irstwoa &	$>$950 &	$>$980 &	$>$6950 &	$>$3700 & SAT	 \\	
\irstwob &	192.0 &	146.9 &	153.6 &	263.0 & 3.6		\\
\irsthreea & 452.1 &	452.0 &	535.5 &	1765.0 & 12\tablenotemark{e}	 \\		
\irsfive &	535.8 &	410.6 &	410.0 &	311.8 & 6 (3.6)\tablenotemark{d}		  \\
\enddata
\tablenotetext{a}{Relative error:  15\%}
\tablenotetext{b}{Relative error:  10\%}
\tablenotetext{c}{Different aperture radii were used to avoid contamination from nearby soruces; see text for discussion.}
\tablenotetext{d}{An aperture radius of 3$\farcs$6 was used for the 8.0~\micron\ image to avoid contamination from variations in the nebular background close to the star.}
\tablenotetext{e}{Since \irsthreea\ does not appear to be a point source (see Sect.~\ref{sect:irsthreea}), a large aperture was used with no correction.}
\label{table:irac_photom}
\end{deluxetable}

\section{Analysis \& Results}
\label{analysis}

\subsection{Spectral Energy Distributions}
\label{sect:SEDs}

\begin{figure}
\epsscale{0.5}
\plotone{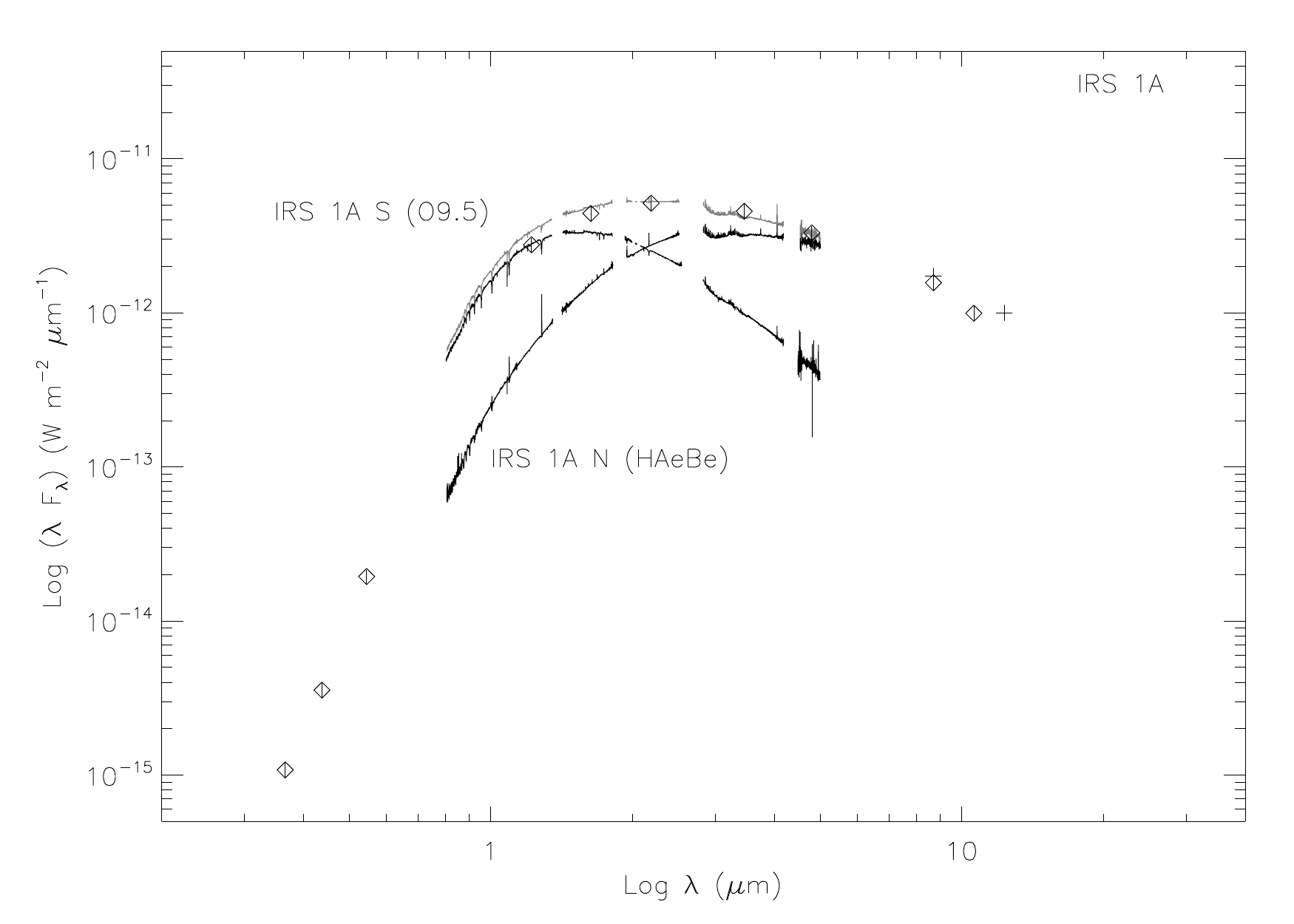}
\caption{Spectral energy distribution for IRS 1a.  Solid line---SpeX near-IR spectra (this study); Grey line---combined near-IR spectrum of 1A North and South;  Crosses---MIRSI mid-IR photometry (this study); Diamonds---\citet{Smith:1985} photometry.  See text for discussion.}
\label{fig:IRS1A_SED}
\end{figure}

In Figures~\ref{fig:SEDs_MS} and \ref{fig:SEDs_YSOs} we plot the full spectral energy distributions for each source from the optical through the mid-IR, including data obtained in this study (SpeX, MIRSI, and Spitzer Archive) and data from \citet{Smith:1985}.  The agreement among the datasets at overlapping wavelengths is generally quite good with two notable exceptions:
\begin{itemize}

\item IRS1A:  The \citet{Smith:1985}, MIRSI, and Spitzer images do not resolve \irsonea\  into its northern and southern components, as seen in our SpeX images (see Section~\ref{sect:spex_obs}).  Comparing the near-IR spectra for N and S to the combined fluxes reported in \citet{Smith:1985} and our MIRSI data, it appears that the longer wavelength emission ($>$4~\micron) is dominated by \irsoneanorth, whereas the shorter wavelengths are dominated by \irsoneasouth\ (Fig.~\ref{fig:IRS1A_SED}).  In the separate SEDs for the northern and southern components, we divide up the fluxes from \citet{Smith:1985} and MIRSI by assigning all data shorter than 1.5~\micron\ to \irsoneasouth, and all points longer than 4~\micron\ to \irsoneanorth.

\item IRS3A:  The absolute levels of the SpeX spectra in the J, H, and K bands do not agree well with either the long wavelength (LXD) SpeX data or the near-IR photometry from \citet{Smith:1985}.  Since the long and short wavelength SpeX data were obtained on the same night, it seems unlikely that this is due to intrinsic variability and instead is due to some issue with the observations.  Indeed the first part of the night during which 3A was observed suffered from variable seeing.   In this case, we decided to scale the J, H, and K band SpeX data by a factor of 0.75 to match the \citet{Smith:1985} fluxes (scaled fluxes for \irsthreea\ are shown in Fig.~\ref{fig:SEDs_MS}).   \irsthreea\ also has notable excess flux in the 8.0~\micron\ IRAC band (channel 4) and the 12.3~\micron\ MIRSI band.  The excess at 8.0~\micron\ is likely due to enhanced emission from the PAH band at 7.7~\micron .  The excess at 12.3~\micron , however, could be due to the 12.7~\micron\ PAH line, crystalline silicates,  12.82~\micron\ [\ion{Ne}{2}] emission, or some combination of all three.  (Since the bandpass for the 12.3~\micron\ filter is only 9.6\%, emission from the strong PAH feature at 11.3~\micron\ would not be included.)    Further mid-IR spectroscopy of \irsthreea\ is required to discern the source of the excesses.
\end{itemize}

Except for \irsfive , all the sources show significant emission in the 5 -- 12~\micron\ range indicative of heated circumstellar dust.    The mid-IR spectral index, $\alpha_{mir} = dlog(\lambda F_{\lambda}) / d\lambda$, for each source was determined by fitting a  line to the $\lambda F_{\lambda}$ values between 3 and 10~\micron.  The results are listed in Table~\ref{table:spectral_features}.

\begin{deluxetable}{lccccc}
\tabletypesize{\scriptsize}
\tablewidth{0pt}
\tablecaption{Spectral Features and Source Properties}
\tablehead{
\colhead{Source}  &   
\colhead{Sp. Type}  &
 \colhead{He I  (1.083~\micron)} & 
\colhead{ $\alpha_{mir}$} &  
 \colhead{CXOW40?\tablenotemark{a}} &  
 \colhead{3.6~cm?\tablenotemark{b}} 
  \\
 &  
 & 
  \colhead{EW (\AA)} &  
 &  
  & 
}
\startdata					
\irsoneanorth & HAe/Be & abs/em  & $-1.3 \pm 0.2$ & Yes? & Yes    \\
\irsoneasouth & O9.5 &   1.81 & -- & Yes? & No  \\
\irsoneb	 & Class II  &  P Cygni  & $-0.6 \pm 0.1$ & Yes & Yes   \\
\irsonec 	 & Class II &  em  & $-0.9 \pm 0.1$ & Yes & Yes     \\
\irstwoa   & HAe/Be  &  P Cygni  & $1.0 \pm 0.1$ & Yes & Yes    \\
\irstwob    & B4  &   1.92 &  $-0.2	\pm 0.3$ & No & No  \\ 
\irsthreea   & B3  &   1.90 & $-0.3 \pm 0.1$ & No & N/A   \\
\irsfive      & B1  &   1.25  & $-1.6	 \pm 0.1$ & Yes & N/A   \\
\enddata
\tablenotetext{a}{\citet{Kuhn:2010}}
\tablenotetext{b}{\citet{Rodriguez:2010}}
\label{table:spectral_features}
\end{deluxetable}

\subsection{Near-IR Spectra:  Main Sequence Stars}
\label{analysis:ms_stars}

\begin{figure}
\epsscale{1.0}
\plotone{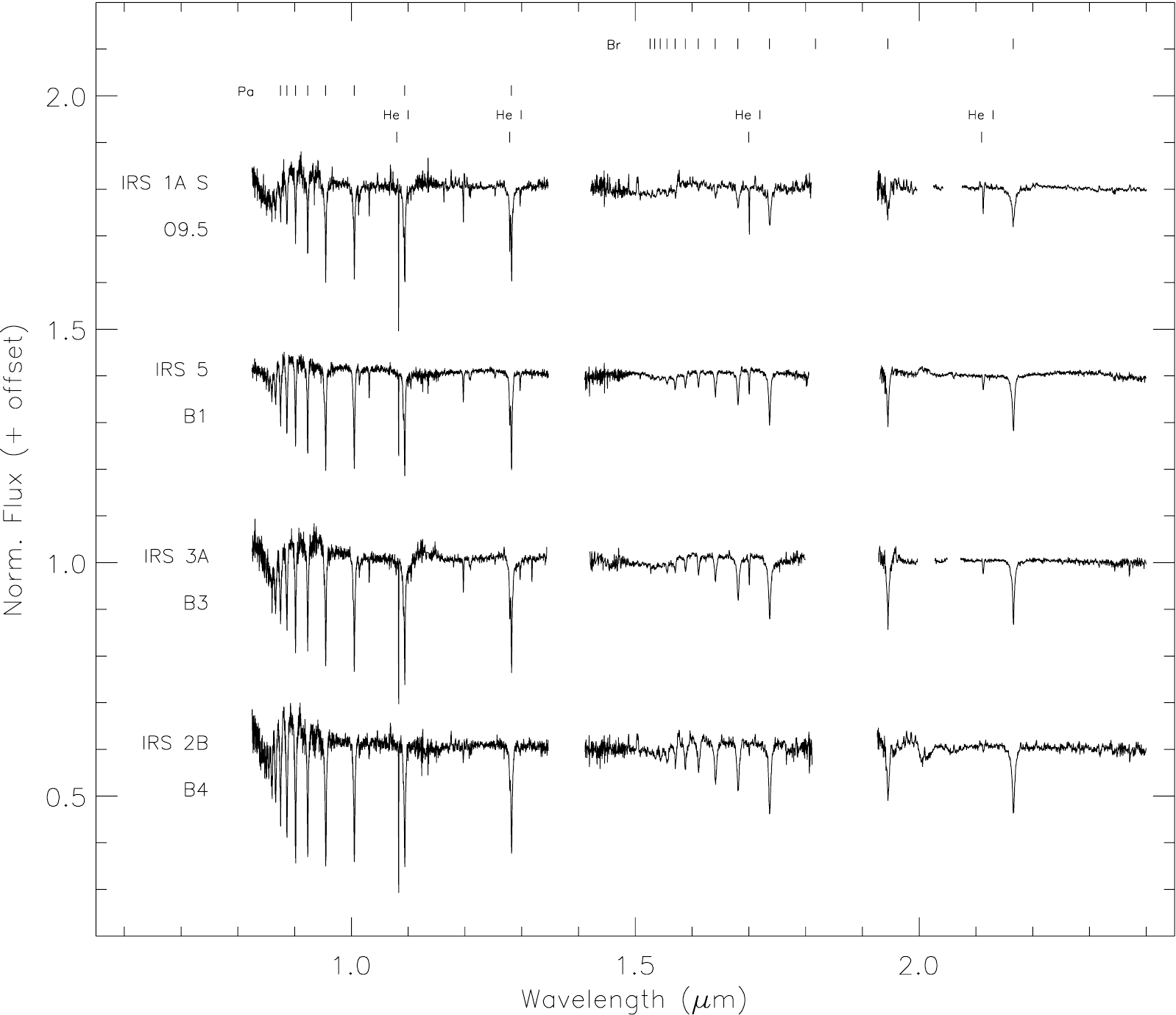} 
\caption{Normalized near-IR spectra for candidate Main Sequence stars in our sample.}
\label{fig:W40_spex_ms}
\end{figure}

Four of the observed sources---\irsoneasouth, 2B, 3A, and 5---have strong Brackett and Paschen series absorption indicative of late-O and early-B type stars (Fig.~\ref{fig:W40_spex_ms}).  None of these sources has obvious \ion{C}{4} emission (2.078~\micron), \ion{N}{3} emission (2.1155~\micron), or \ion{He}{2} absorption (2.1885~\micron) which restricts the spectral types to O9 or later.  Determining spectral types based on line equivalent widths alone is notoriously difficult~\citep{Hanson:2005}, so we have compared our spectra directly with early-type stars from the IRTF spectral library~(Vacca et.~al., in preparation). Derived spectral types are listed in Table~\ref{table:spectral_features} and are accurate to about +/- 1 sub-type.  We also measured the equivalent widths of \ion{He}{1} (2.1126~\micron) and Br$\gamma$ (2.16~\micron) lines for each source and compared them to the \citet{Hanson:1996} and \citet{Blum:1997} spectral atlases.  In all cases, our EWs are consistent with early-B spectral types, but note that it is very difficult to discriminate between subtypes based on EWs alone.  The existence of O and B main sequence stars at the center of the W40 IR cluster implies an upper limit on the age of the W40 cluster of $\approx 7$~My.

All of the main sequence sources in our sample exhibit deep, unresolved \ion{He}{1} absorption at 1.083~\micron\ (see Table~\ref{table:ms_stars}), which is often the deepest line in the spectrum.  Though relatively rare for ``normal'' O and B-type stars~\citep{Conti:1999}, it is quite prominent in both emission and absorption for massive stars with dense winds including some Be stars~\citep{Groh:2007}.  Since the line widths are very narrow, and none of our main sequence (MS) stars exhibit Be star properties, it seems unlikely that the strong \ion{He}{1} absorption is due to stellar winds.  Note also that none of our MS stars, except \irsfive , is a strong X-ray emitter, which is also a signature of stellar winds in B stars~\citep{Cassinelli:1994}.  

Relatively strong \ion{He}{1}~1.083~\micron\ absorption can also be produced by non-LTE effects in the atmospheres of early-type stars~\citep{Lennon:1989,Przybilla:2005}.  However, the equivalent widths we measure for the W40 sources are far larger than those measured by \citet{Lennon:1989}, or derived from the models of \citet{Przybilla:2005}.  In fact, the 1.083~\micron\ absorption is {\em also} far stronger than that observed in the reference spectra of the IRTF spectral library that were used to determine the spectral types of the W40 sources.  These results strongly suggest that the 1.083~\micron\ absorption is not due to non-LTE effects in the stellar atmospheres of these sources.  \ion{He}{1} absorption is also thought to be associated with coronal magnetic activity,  and hence is commonly observed for low-mass main sequence stars and close binaries~\citep{Zirin:1982}.   It seems likely that all of our MS sources are in fact  binaries with close and/or low-mass companions.  In addition,  \irsfive\ is a known X-ray source~\citep{Kuhn:2010} which further supports this conclusion for that source.  

For each main sequence star in our small sample, we simultaneously determined the extinction ($A_V$) and distance by fitting a reddened  atmospheric model to the SpeX near-IR spectra using a $\chi^2$ minimization routine.  We adopted a  Kurucz standard model atmosphere~\citep{Kurucz:1993} for each target with the stellar radius and effective temperature for each spectral type taken from  \citet{deJager:1987}. We adopted the reddening law from \citet{Fitzpatrick:2007} for a fixed $R_V$ of 3.1.  The results, including our adopted radii and and effective temperatures, are presented in Table~\ref{table:ms_stars} with resulting reddened model atmospheres plotted in Figure~\ref{fig:SEDs_MS}.   In general, the fits to the near-IR data are excellent; the values for distance and $A_V$ agree with previous studies quite well.  Note that the reddened model SED for both \irsoneasouth\ and \irsthreea\ is always well below the UBV photometry from \citet{Smith:1985}, possibly due to uncertainty in the optical-to-IR extinction law for W40.  

\begin{deluxetable}{lccccccccc}
\tabletypesize{\scriptsize}
\tablewidth{0pt}
\tablecaption{Summary of Near-IR Atmospheric Model Fits to Main Sequence Sources}
\tablehead{
\colhead{Source}  &  \colhead{Sp. Type}  &  \colhead{$T_{eff}$\tablenotemark{a}} &
\colhead{Radius\tablenotemark{a}} & \colhead{Model $T_{eff}$}  & \colhead{log($Q_0$)} &
\colhead{$E(B-V)$\tablenotemark{b}} & \colhead{$A_V$} & 
\colhead{D} & \colhead{D - range}    \\
  &  &  \colhead{(K)} &
\colhead{($R_{\sun}$)} & \colhead{(K)}  & &
  &  & 
\colhead{(pc)} & \colhead{(pc)}  
}
\startdata					
\irsoneasouth & O9.5 & 32210 & 7.71 & 32000 & 47.8 & 3.4 & 10.6 & 536.0 & 441---578  \\
\irstwob & B4 & 16904 & 3.43 & 17000 & 42.8 & 2.9 & 9.1 & 455.5 & 396---526  \\ 
\irsthreea & B3 & 18793 & 3.76 & 19000	 & 43.7 & 3.0 & 9.2 & 	349.0 & 312---416  \\
\irsfive & B1 & 25410 & 5.16 & 25000 & 45.7 & 2.8 & 8.5 & 469.1 & 	340---686    \\
\enddata
\tablenotetext{a}{\citet{deJager:1987}}
\tablenotetext{b}{All fits assume $R_V = 3.1$. }
\label{table:ms_stars}
\end{deluxetable}

The dominant source of error in the fits is the $\pm$1 subtype uncertainty in the spectral type whereas the dominant source of error in $A_V$ is the extinction law.  To estimate the uncertainties (particularly for the distance), we ran the fitting routine for +/- 1 spectral-subtype (with an associated change in $R$ and $T_{eff}$) for each source with the resulting range in distance shown in Table~\ref{table:ms_stars}.  For \irsoneasouth , we also tried $R$ and $T_{eff}$ values from \citet{Vacca:1996} and \citet{Martins:2005}.  The \citet{Vacca:1996} values for $R$ and $T_{eff}$ are somewhat larger than those of \citet{deJager:1987}, which pushes \irsoneasouth\ about 100~pc further away.  The value for $T_{eff}$ from  \citet{Martins:2005} is roughly the same as that of  \citet{deJager:1987}, but Martins~et~al. give a much smaller radius for an O9.5 star, which brings \irsoneasouth\ about 40~pc {\em closer}.  

Although the acceptable ranges in distance for each source overlap very nicely, there are some discrepancies among the best-fit values derived for each source.  The agreement in distance between \irstwob\ and 5 is striking ($< 3$\%, well within the estimated errors).  The large error in distance for \irsfive\ is due to the large spread in effective temperature for early B-stars.  \irsthreea\ appears to be much closer than the others which does not seem physically plausible given the nearly identical $A_V$ and assumption of cluster membership.  In addition, the model fit for this source is not as good as the fits for the other sources.  This may be due to a possible close binary companion, as suggested above (\irsthreea\ will be discussed further in Section~\ref{sect:irsthreea}).  On the other hand, \irsoneasouth\  appears to be somewhat farther than both \irstwob\ and 5.  This could be rectified if 1A~South were one subtype later and \irstwob\ and 5 roughly one subtype earlier.  An alternative explanation is that the distance to 1A~South is correct, and that both \irstwob\ and 5 are {\em overluminous}.  This would further support the argument that these two objects are in fact close binaries, as indicated by the strong \ion{He}{1} 1.083~\micron\ absorption.  Changes in the assumed extinction law do not affect the distance strongly, but can produce different values for $A_V$.  Changing the spectral type by one sub-type (as above) generally causes a variation in $A_V$ of less than a few percent.  Adopting a value of $R_V$ of 5.0 causes a decrease in the $A_V$ of roughly 20\%.

Except for \irsoneasouth\ (which is not resolved from 1A~North in the mid-IR), all of the main sequence objects show at least some mid-IR excess indicative of heated circumstellar dust.  Based on the combined fluxes (Section~\ref{sect:SEDs}), we have attributed the majority of the mid-IR flux from \irsonea\ to the northern component.  It is, however, possible that some of the mid-IR flux is due to circumstellar dust associated with \irsoneasouth.  For \irstwob, the IR excess does not become substantial until $\approx 6$~\micron , and is very strong at 10~\micron.  It is not clear whether the flux at 10~\micron\ is due to silicate or thermal emission;  further observations at longer wavelengths are required.  In either case, a flux deficit in the near-IR with strong IR emission at longer wavelengths is often modelled by an optically thick disk with a central``hole''~\citep{Calvet:2005,DAlessio:2005}, which may be created by photoevaporation~\citep{Clarke:2001} or the presence of a close companion~\citep{Uchida:2004,Rice:2003}, or both.  On the other hand, the mid-IR excess for \irsfive\ decreases uniformly, suggesting a much cooler, older disk.  The mid-IR excess for \irsthreea\ is very flat from 3 to 10~\micron\ punctuated by excess emission in the 8 and 12~\micron\ bands;  this source will be discussed further in Section~\ref{sect:irsthreea}.   The presence of circumstellar dust for these sources indicates that there has been insufficient time for these stars to dissipate their natal disks, and hence they may have only recently arrived on the main sequence.

\subsection{Near-IR Spectra:  Young Stellar Objects}

\begin{figure}
\epsscale{1.0}
\plotone{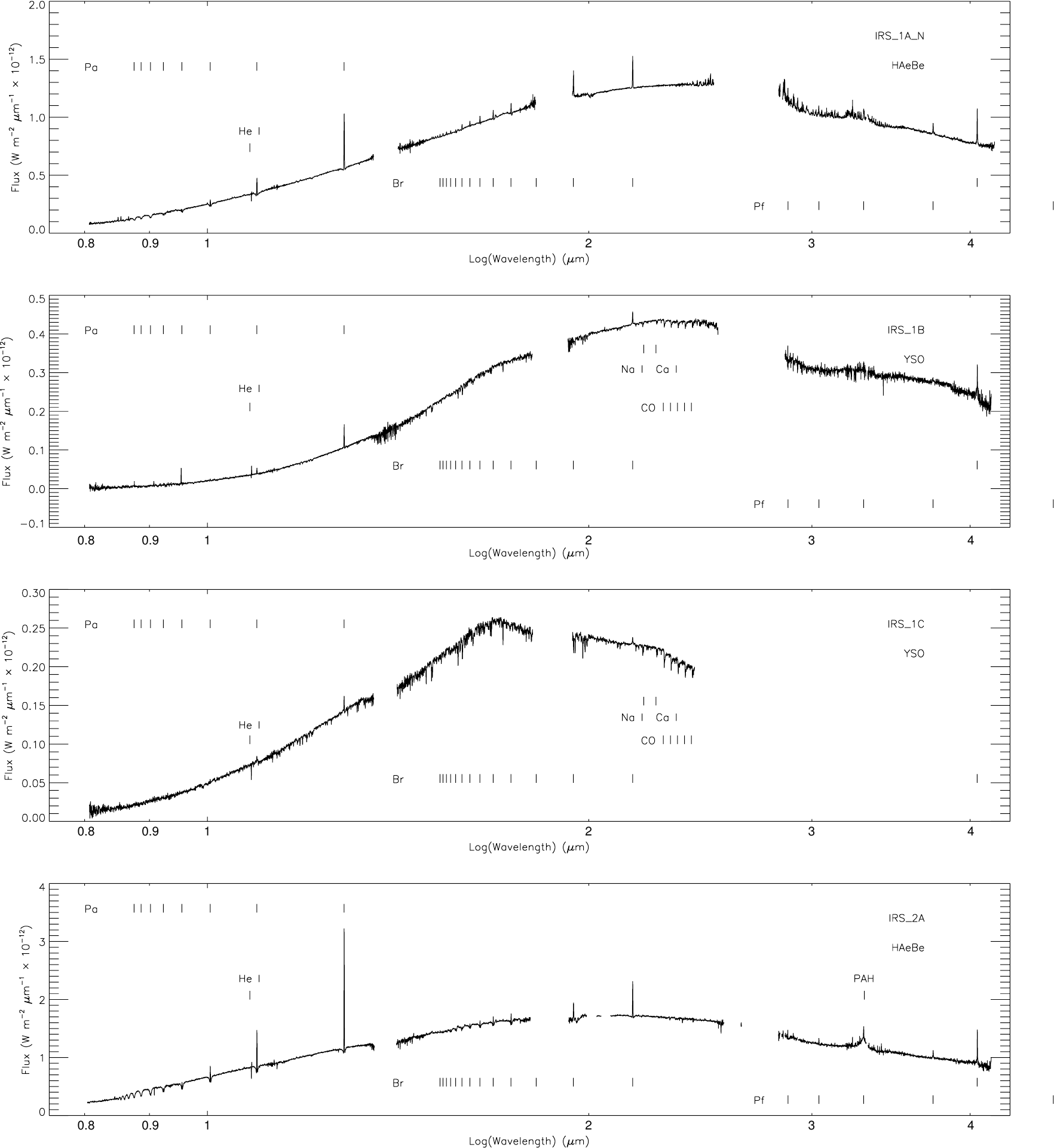} 
\caption{Final SpeX flux-calibrated spectra:  Pre-Main Sequence and Young Stellar Objects.}
\label{fig:W40_spex_yso}
\end{figure}

The sources \irsoneanorth, \irsoneb, \irsonec, and \irstwoa\ all have very different near-IR spectral features compared to the Main Sequence stars identified in our sample (see Fig.~\ref{fig:W40_spex_yso} and Table~\ref{table:spectral_features}).   Both \irsoneanorth\ and \irstwoa\ have strong Brackett and Paschen series emission, coupled with strong absorption in the higher level Paschen lines.  Both stars show \ion{Ca}{2} and \ion{O}{1} in emission as well.  Based on these spectral features, and the strong mid-IR emission for both, we conclude that \irsoneanorth\ and \irstwoa\ are likely heavily reddened Herbig~Be stars.  Using the strongest hydrogen emission lines in both sources, we derive an $A_V$ of $\approx 10$, assuming Case~B recombination, which agrees with that derived from the main sequence continuum fitting.   All of the YSOs in our sample exhibit strong 1.083~\micron\  emission and absorption (including P Cygni profiles for \irsoneb\ and \irstwoa) which is common in both T Tauri and HAeBe stars~\citep{Edwards:1987}.

The near-IR spectra for \irsoneb\ and 1C are generally smooth except for modest Brackett and Paschen series emission coupled with strong \ion{Na}{1}, \ion{Ca}{1}, and CO absorption.  The near-IR spectral features and the mid-IR spectral index ($\alpha_{mir}$) suggest that both of these sources are heavily reddened Class~II low mass Young Stellar Objects (YSOs)\citep{Greene:1996}.

\clearpage

\section{Discussion} 
\label{discussion}

\subsection{The Distance to W40}
\label{distance}

Depending on the technique used, the distance to the W40 complex has been estimated to be between 300 and 900~pc~\citep[and references therein]{Rodney:2008,Kuhn:2010}.  If we exclude \irsthreea\ (which has a relatively poor fit, and is also exceptional in other ways, see Section~\ref{sect:irsthreea}), then the best estimate for the distance based on our results (for 3 stars) is between 455~pc and 536~pc.  We can confidently rule out distances less than 340~pc and greater than 686~pc; making it very unlikely that W40 is part of the same complex that includes the Serpens star forming regions, as assumed by~\citet{Bontemps:2010}.  Our best estimates are somewhat lower than the distance estimate of 600~pc derived from X-ray luminosity function (XLF) fitting method used by \citet{Kuhn:2010}. (In fact, a distance of 500~pc produces a qualitatively worse fit to the XLF according to the authors.) Our distance determination has the advantage of being based on (admittedly only a few) well-determined spectral types, but only for a few sources that happen to be OB stars with a wide range of possible luminosities.  The advantage of the XLF method is that it is based on statistics over a large number of observed sources;  the disadvantage is that one must assume a ``universal XLF'' that may not apply to all star forming regions.  For the remainder of the paper, we will adopt a distance of 500~pc.

\subsection{IRS 3A}
\label{sect:irsthreea}
As indicated in Section~\ref{sect:SEDs}, \irsthreea\ has significant mid-IR excess with a generally flat spectral index ($\alpha_{mir} =  -0.3$).   In addition,  \irsthreea\ is also {\em resolved} as a circular source at longer mid-IR wavelengths.  This can be seen readily in both the MIRSI 8.7 and 12.3~\micron\ images (FIg.~\ref{fig:W40_MIRSI}), and in the IRAC channel 4 images (see red channel in Fig.~\ref{fig:W40_IRAC_134}).  In fact, \irsthreea\ is point-like only in the IRAC channel 1 (3.6~\micron) data;  in the other bands it has a roughly Gaussian profile with a FWHM that exceeds that of the   corresponding PSF.  The measured FWHM as a function of wavelength (with comparison to the PSF FWHM) is shown in Table~\ref{table:irsthreea_size}. 
At 500~pc, the measured FWHM at 4.5~\micron\ corresponds to a size of 1050~AU, increasing to 2700~AU at 12.3~\micron.  The fact that \irsthreea\ is resolved in the near- to mid-IR indicates that the emission must be coming from a circularly symmetric distribution of dust---either a spherical envelope or a disk viewed roughly face-on--- with a radius of roughly 500~AU.  There are, however, issues with both of these interpretations.  In general, near- to mid-IR emission from circumstellar disks around intermediate mass stars (e.g. HAeBe stars) is generated at the disk surface where dust temperatures exceed 250~K.  For most flaring disk models, this corresponds to regions at the disk surface within 10~AU of the central star~\citep[e.g.][]{Dullemond:2002}, and hence would be {\em unresolved} in our observations.  On the other hand, an optically thick spherical envelope would produce additional extinction in the optical that would produce a larger $A_V$ for \irsthreea\ than the other stars in W40, which is not observed.  A more probable explanation is that the resolved emission from 4 -- 12 \micron\ is due to emission from small grains (transiently heated by the strong UV emission from \irsthreea) in an optically thin, dusty nebula (or shell) immediately surrounding the star~\citep[e.g.][]{Hartmann:1993}.  If this is the case, then PAH and [\ion{Ne}{2}] emission from this dusty nebula could also explain the excess flux observed in the 8.0 and 12.3~\micron\ bands (see Figure~\ref{fig:SEDs_MS}).  This dusty shell may be a relic from an earlier pre-main sequence (Herbig~Be) phase which has not yet been dispersed by the stellar winds of \irsthreea .


\begin{deluxetable}{lccc}
\tabletypesize{\footnotesize}
\tablewidth{0pt}
\tablecaption{IRS~3A:  Size vs. Wavelength}
\tablehead{
\colhead{Waveband}  &  
\colhead{PSF\tablenotemark{a} }  &  
\colhead{IRS 3A}  &
\colhead{IRS 3A}  \\

\colhead{(\micron)}  &  
\colhead{FWHM (\arcsec)}  &  
\colhead{FWHM (\arcsec)}  &
\colhead{FWHM (AU)\tablenotemark{b}}
}
\startdata					
IRAC1, 3.6 &	1.66 &	1.764 &	882 \\
IRAC2, 4.5 &	1.72 &	2.1 &	1050  \\
IRAC3, 5.8 &	1.88 &	2.58 &	1290 \\
IRAC4, 8.0 &	1.98 & 4.158 &	2079 \\
MIRSI, 8.7	 & 1.21 &	4.32 &	2160 \\
MIRSI, 12.3 &	1.32  &	5.4 &	2700 \\
\enddata
\tablenotetext{a}{IRAC PSF taken from mean FWHM of the point response function (PRF) reported in the IRAC Instrument Handbook.  MIRSI PSF based on measured performance and is nearly diffraction limited.}
\tablenotetext{b}{Assuming $d = 500$~pc.}
\label{table:irsthreea_size}
\end{deluxetable}

The fact that the Kurucz model fits yield a distance to W40 that is far closer than that of the other MS stars in our (admittedly small) sample, and the presence of strong \ion{He}{1} absorption strongly suggests that \irsthreea\ is in fact a close binary.  Increasing the distance of \irsthreea\ by a factor $\approx$1.3 would place the binary at the same distance as \irstwob\ and \irsfive ,  and would result in a  $\approx$1.7 fold increase in the luminosity, which could be accounted for by a companion of mid-B spectral type. It is also possible that \irsthreea\ is overluminous because it is a ``bloated'' star, i.e. the stellar radius is much larger than a typical MS B3~\citep[e.g][]{Ochsendorf:2011}.  Pre-main sequence evolutionary tracks for intermediate mass stars show a significant increase in total luminosity just before the star settles onto the main sequence~\citep{Palla:1993,Hosokawa:2009}.  It is possible that \irsthreea\ is nearing (or at) this peak in luminosity, which would contribute to the poor stellar model fit, result in a shortened distance, and explain the existence of the extended dusty envelope.  In this case, the strong \ion{He}{1} absorption could still be due to a much lower mass binary companion.

\subsection{Compact Radio Emission}

All of the sources we identify as YSOs in our sample are also detected in the radio survey of~\citet{Rodriguez:2010} (see Table~\ref{table:spectral_features}).  Only two of the main sequence stars in our study are also included by the radio maps of~\citet{Rodriguez:2010} (\irsoneasouth, and \irstwob) and neither is detected at 3.6~cm.  \citet{Rodriguez:2010} suggest that the non-variable compact radio sources are likely due to ultra-compact \ion{H}{2} (UC\ion{H}{2}) regions around young, high mass stars.  This seems unlikely, however, as the size of the unresolved radio sources at a distance of 500~pc would be less than $\approx 100$~AU---much smaller than a typical UC\ion{H}{2} region\citep{Kurtz:2005}.  It seems more likely that the 3.6~cm continuum flux would be due to free-free emission from shocked gas within 100~AU of the YSO caused by a jet or outflow.  It could also be driven by synchrotron emission associated with the stellar magnetic field, a conclusion that is  supported by the fact that all the YSOs in our sample are also strong X-ray sources (see Table~\ref{table:spectral_features}).

\subsection{The W40 \ion{H}{2} Region}

The predicted Lyman continuum (LyC) fluxes for each of our main sequence sources (based on the \citet{Kurucz:1993} model) is given in Table~\ref{table:ms_stars}.  The total LyC flux in the region is dominated by  the O9.5 star \irsoneasouth\ at 6.3$\times 10^{47}$~photons~s$^{-1}$.  Based on our spectral type analysis (Sect~\ref{analysis:ms_stars}), it is unlikely that \irsoneasouth\ is earlier than O9 which places an upper limit on the LyC flux available to drive the \ion{H}{2} region of $\approx 1.6 \times 10^{48}$ photons~s$^{-1}$.  This upper limit agrees very nicely with the derived LyC flux of $\sim1.5\times10^{48}$ photons~s$^{-1}$ from radio observations~\citep{Goss:1970vn,Smith:1985}.  Based on the IR luminosity, \citet{Smith:1985} originally suggested that \irstwoa\ must be the dominant energy source in W40.  We conclude, however, that  \irsoneasouth\ must be the main source of ionizing radiation.  

\begin{figure}
\epsscale{0.5}
\plotone{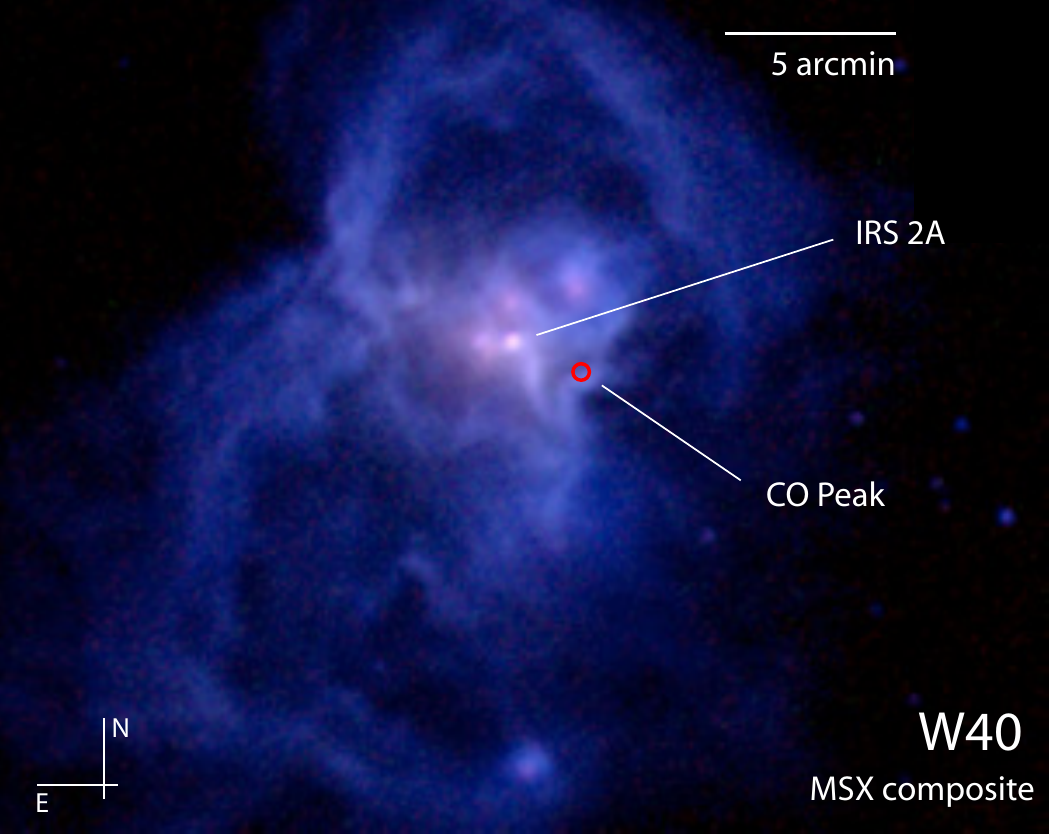}
\caption{MSX 3-color composite of W40 at 8.3 (blue), 12.1 (green), and 14.6~\micron\ (red).  Field of view is approximately 30\arcmin.  Red circle marks the center of the peak CO emission \citep{Zhu:2006fu}. }
\label{fig:W40_MSX}
\end{figure}

The morphoplogy of the \ion{H}{2} region in the thermal IR is quite different from that seen in the POSS plates and 2MASS images.  Images of the region from the Mid-course Space Experiment (MSX) obtained from the NASA \anchor{http://irsa.ipac.caltech.edu/Missions/msx.html}{Infrared Science Archive (IRSA)}\footnote{
\url{http://irsa.ipac.caltech.edu/Missions/msx.html}}
 (see Fig.~\ref{fig:W40_MSX}) reveal an hour-glass shaped structure centered roughly on \irstwoa, approximately 28\arcmin\ in height and about 17\arcmin\ in width~\citep{Rodney:2008}, corresponding to 4.1 by 2.5~pc at a distance of 500~pc.   The long axis of the hourglass is at a PA of $\sim 300$~degrees (NNW).  The structure is most prominent in the MSX A-band (8.3 ~\micron) images, where the brightness is likely dominated by emission in the PAH line at 7.7~\micron. The hourglass does not appear related to the weak CO outflow centered on the dense material roughly 3\arcmin\ to the West~\citep{Zhu:2006fu}.  We interpret this double-lobed morphology as a pinched waist ``bubble''  blown by a symmetric wind, most likely from the O9.5 star \irsoneasouth.  The waist axis of the hourglass lies along a PA of roughly 60~degrees (ENE).  The Spitzer IRAC images (Fig.~\ref{fig:W40_IRAC_134}) reveal numerous bright-rimmed sheets and clumps (``elephant trunks'') along this axis, nearly all of which point back toward \irsonea, further underscoring \irsoneasouth\ as the dominant wind and UV source of the region.   Far-IR and mm-wave observations of W40 reveal 36 deeply embedded sources (prestellar cores, and Class 0,I objects) indicating that star formation is on-going, and has perhaps been initiated by the pressure wave associated with the wind-bubble~\citep{Maury:2011}.

\clearpage

\section{Summary}
\label{summary}

We have presented the first detailed analysis of the eight brightest members of the galactic \ion{H}{2} region W40 based on medium resolution near-IR spectroscopy (IRTF/SpeX) and mid-IR imaging (IRTF/MIRSI, Spitzer IRAC).  SpeX guide camera images reveal that \irsonea\ is in fact a binary (with ``North'' and ``South'' components).  In addition, \irsoned\ breaks up into at least 7 distinct sources in the near-IR.  Our MIRSI images reveal 5 new sources in the mid-IR (IRS~1 E and F, IRS~2 D -- F), two of which are diffuse in nature.  

Using the SpeX near-IR spectra we have classified all 8 of the sources observed based on near-IR emission and absorption features (see Table~\ref{table:spectral_features}).  Four of the objects appear to be late-O or early-B stars.  Two of the stars have spectral features indicative of Herbig Ae/Be stars, and the remaining two appear to be low mass YSOs (Class~II).  Strong \ion{He}{1} absorption at 1.083~\micron\ strongly suggests that the main sequence objects may in fact be close binaries.  Spectral energy distributions reveal that 7 of the 8 sources have mid-IR excesses indicative of circumstellar disks.  

For the main sequence sources, we simultaneously derived $A_V$ and $D$ for each star by fitting a reddened stellar atmosphere appropriate to the spectral type.  In general, the agreement between the sources is good.  We find a best estimate for the distance of  455---535~pc with $A_V$ ranging from 8.5 to 10.6.  Our derived distance agrees well with previous studies, though is somewhat lower than the recent determination by ~\citet{Kuhn:2010} based on XLF fitting.  Our distance determination indicates that the radio emission from some of the sources reported in \citet{Rodriguez:2010} cannot be due to UC\ion{H}{2} regions (due to size constraintss) and is instead most likely due to shocked gas due to jets and/or outflows from the source.  

\irsthreea\ is found to be somewhat peculiar in a few ways:  first, despite having a fairly clear B3 spectral type, stellar model fits were quite poor and the distance to the source was found to be significantly smaller than that for the 3 other MS stars.  Furthermore, \irsthreea\ appears to have a substantial dusty envelope, at least 2700~AU in size, that is resolved at mid-IR wavelengths.  We suggest that \irsthreea\ is either a close binary composed of two early-B stars, or a PMS star with a close companion and/or a strong stellar wind.  

Our observations indicate that the O9.5 star  \irsoneasouth\ is the dominatant source of LyC photons for the W40 \ion{H}{2} region (as opposed to \irstwoa , as originally suggested by \citet{Smith:1985}). Furthermore, \irsoneasouth\ is situated at the center of the pinched waist morphology seen in wide-field mid-IR images (MSX), suggesting that it is the source of the stellar wind that has blown this bubble over time.

\acknowledgements
This research has been supported in part by the Universities Space Research Assoc. (USRA) under contract 209000771 to R. Y. Shuping at the Space Science Institute. Data was obtained at the Infrared Telescope Facility, which is operated by the University of Hawaii under Cooperative Agreement NNX-08AE38A with the National Aeronautics and Space Administration, Science Mission Directorate, Planetary Astronomy Program.  This work is based in part on observations made with the Spitzer Space Telescope, and data obtained from the NASA/ IPAC Infrared Science Archive, both of which are operated by the Jet Propulsion Laboratory, California Institute of Technology under a contract with the National Aeronautics and Space Administration.

{\it Facilities:} 
\facility{IRTF (SpeX, MIRSI)}; 
\facility{Spitzer (IRAC)};
\facility{MSX};
\facility{IPAC (IRSA,MOPEX)}.

\clearpage



\end{document}